\documentclass[journal]{rmaa}

\suppressfulladdresses 
\usepackage{psfrag,color}
\usepackage{booktabs}
\usepackage{longtable}
\usepackage{natbib}
\usepackage{paralist}
\usepackage{amsmath}
\usepackage{tabularx}
\usepackage[flushleft]{threeparttable}
\usepackage{graphicx}

\suppressfulladdresses 
\usepackage{psfrag,color}
\usepackage{booktabs}
\usepackage{longtable}
\usepackage{natbib}
\usepackage{paralist}
\usepackage{amsmath}
\usepackage{tabularx}
\usepackage[flushleft]{threeparttable}
\usepackage{graphicx}
\usepackage{color,ulem}
\usepackage{array}
\usepackage{tablefootnote}
\usepackage{lineno}
\usepackage{pdflscape} 

\usepackage{verbatim} 
\newcommand{\kms}{km\,s$^{-1}$}
 
\SetYear{2024} \SetConfTitle{RevMexAA}

\title{Dense Molecular Gas and dusty torus in NGC\,4303}

\author{ Ángel~A. Soní\altaffilmark{1},
 Irene Cruz-González\altaffilmark{1},
Martín Herrera-Endoqui\altaffilmark{1,2}, \\
 Erika Benítez\altaffilmark{1},   Yair Krongold\altaffilmark{1}
 \& Arturo~I. Gómez-Ruiz\altaffilmark{3} }

\altaffiltext{1}{Instituto de Astronom\'{\i}a, Circuito Exterior, C. U., Universidad Nacional Aut\'onoma de M\'exico, 04510 CdMx, México. }
\altaffiltext{2}{Estancia Posdoctoral por M\'exico, CONAHCYT, Coordinaci\'on de Apoyos a Becarios e Investigadores.}
\altaffiltext{3}{CONAHCYT-Instituto Nacional de Astrof\'isica \'Optica y Electr\'onica, Luis E. Erro 1, 72840 Tonantzintla, Puebla, México.}

\shortauthor{Soní et al.}

\shorttitle{Dense Molecular Gas and Dusty Torus in NGC\,4303}

\fulladdresses{
\item I.Cruz-Gonz\'alez: Instituto de Astronom\'{\i}a, Universidad Nacional Aut\'onoma de M\'exico, Apdo. Postal 70-264, Cd. de M\'exico (irene@astro.unam.mx).}

\listofauthors{A.~A. Soní, I. Cruz-González, M. Herrera-Endoqui, E. Benítez1,
Y. Krongold & A.~I. Gómez-Ruiz}
\indexauthor{Soní, A.~A.}
\indexauthor{Cruz-González, I.}
\indexauthor{Herrera-Endoqui, M.} 
\indexauthor{Benítez, E.} 
\indexauthor{Kromgold, Y.} 
\indexauthor{Gómez-Ruiz, A.~I.}

\abstract{Spectrum analysis at 3 mm of the central region ($r\sim$800 pc) of NGC\,4303 showed molecular gas lines of both dense gas tracers (HCN, HNC, HCO$^+$, and C$_2$H) and diffuse gases  ($^{13}$CO and C$^{18}$O). 
Molecular gas derived parameters: $H_2$ mass
$M_{H_2}$=(1.75$\pm$0.32)$\times10^{8}$ M$_{\odot}$; radial velocity  
V$_{dense}=$178$\pm$60 km\,s$^{-1}$, and V$_{CO}=$151$\pm$29 km\,s$^{-1}$; 
HCN luminosity $L_{HCN}$=(7.38$\pm$1.40)$\times10^{6}\,\,K\,\,km\,\,s^{-1}\,pc^{2}$; dense gas mass $M_{dense}$=(4.7$\pm$0.3) $\times 10^{7}$ M$_{\odot}$, and dense gas tracers abundances indicating that dense gas contributes significantly to the total molecular gas mass. 
To explore the AGN nature and
central dusty torus of the galaxy, CIGALE was used to fit the integrated spectral energy distribution from submillimeter to  UV frequencies.
Large torus properties are estimated: 
luminosity $L_{TORUS}$\,=\,(7.1$\pm$2.8) $\times 10^{43}$ erg s$^{-1}$ and line of sight inclination of 67$\pm$16$^\circ$, which is consistent with a Type 2 AGN; total infrared luminosity $L_{IR}$=(3.51$\,\pm$\,0.30)$\times 10^{44}$ erg s$^{-1}$; star formation rate $SFR$=6.0$\pm$0.3 M$_\sun$\,yr$^{-1}$; and found that the AGN contribution is marginal at $\sim$20\%.}

\resumen{El análisis del espectro  a 3 mm de la región central de NGC\,4303 en un radio de $\sim$800 pc, mostró líneas de gas molecular de trazadores de gas denso HCN, HNC, HCO$^+$ y C$_2$H, y de gas difuso $^{13}$CO y C$^{18}$O. Se
obtuvieron parámetros del gas molecular:
masa del hidrógeno molecular total de la región central,  $M_{H_2}$=(1.75$\,\pm\,$0.32)$\times10^{8}$ M$_{\odot}$; velocidad radial 
de gas denso $V_{dense}=$178$\,\pm\,$60 km\,s$^{-1}$, y gas difuso $V_{CO}=$151$\,\pm\,$29 km\,s$^{-1}$
luminosidad de HCN $L_{HCN}$=(7.38$\,\pm\,$1.40)$\times10^{6}\,\,K\,\,km\,\,s^{-1}\,pc^{2}$,  masa de gas denso $M_{dense}$=(4.7$\,\pm\,$0.3)$\times 10^{7}$ M$_{\odot}$, lo que muestra que el gas denso tiene una contribución  significativa a la masa total del gas molecular. Las abundancias de trazadores de gas denso fueron derivados para explorar los procesos físicos que prevalecen en la región nuclear de NGC\,4303.  
Para explorar la naturaleza activa y el toro de polvo central de la galaxia se utilizó CIGALE,
para ajustar la distribución espectral de energía integrada de sub-mm al UV. Se estimaron propiedades del toro, como luminosidad $L_{TORUS}$\,=\, (7.1$\pm$2.8)$\times 10^{43}$ erg s$^{-1}$ y  ángulo de visión 67$\pm$16$^\circ$, la cual es consistente con un AGN tipo 2; luminosidad total infrarroja  $L_{IR}$=(3.5$\pm$0.3)$\times 10^{44}$ erg s$^{-1}$; tasa de formación estelar $\ SFR$=6.0\,$\pm$\,0.3 M$_\sun$\,yr$^{-1}$, y se encontró que la contribución del AGN es marginal $\sim$20\%.
}
\addkeyword{Galaxies: active} 
\addkeyword{Galaxies: ISM}  
\addkeyword{Galaxies: individual (NGC\,4303)} 
\addkeyword{Galaxies: molecular gas}

\begin{document}
\maketitle

\section{Introduction}
\label{sec:intro}

Molecular gas characterization of the few central kiloparsecs region of galaxies  is important to understand how the gas  is accreted into the supermassive black hole (SMBH), and  {the role of the last one} in the host galaxy evolution \citep[e.g.][]{2011A&A...528A..30C, 2019ApJ...880..127J}. {The activity in the nuclei of galaxies (AGN) can be dominated by SMBH accretion or starburst (SB) processes, which have different physical and chemical properties that affect their interstellar medium (ISM). 

{ In some AGN the obscuration of the nuclear region by dust and gas  is parameterized by the neutral hydrogen column density $N_H$}. Infrared and  
sub-millimeter observations are less susceptible to dust, especially in the 3 mm window where the brightest emission lines such as HCN or HCO$^+$ trace dense molecular gas, whereas the $^{12}$CO line traces diffuse gas. {These lines have proven useful for understanding the central ISM in nearby galaxies \citep[for example,][]{2011A&A...528A..30C,2015A&A...579A.101A}. On the other hand, in the infrared the dust emission can be observed and studied using broad-band photometry. }

Simultaneous observations of $^{12}$CO isotopic varieties $^{13}$CO and C$^{18}$O of galaxies are valuable for understanding the relationship between the  {bulk of} molecular gas and its environment influenced by the SMBH accretion \citep[e.g.,][]{2016IAUS..315..207G}. In addition, these lines are crucial for deriving the  {gas mass} that will be converted into stars. 

On the other hand, observations of molecular dense gas tracers such as HCN, HNC, HCO$^+$, CS, and N$_2$H are valuable because their critical densities are of the same order as the molecular cloud cores, which are the star formation sites. Such studies base their analysis on {the emission line ratios, for example, HCN/HCO$^+$ has been used as a discriminator between AGNs, segregating them into non-Seyfert galaxies (nuclear SB) or pure Seyfert (Sy) galaxies} \citep[][]{2003ASPC..289..349K}. Additionally, HCN/HCO$^+$  {has been used as an indicator of the presence of AGNs, although this has been challenged several times without conclusive evidence \citep[e.g.,][]{2011A&A...528A..30C, 2020ApJ...893..149P}}. {HCN/HNC has been used to characterize photon-dominated regions (PDRs) and X-ray dominated regions (XDRs) \citep[e.g.,][]{2007A&A...461..793M}. In addition, both ratios} show variations between different galaxy locations such as the center, disk, spiral arms and interarm regions in different studies \citep[e.g.,][]{2016ApJ...822L..26B,2019ApJ...880..127J,2019COMINGS,2020PASJ...72...90M}. Finally, the dense to diffuse gas ratio HCN/$^{12}$CO associated to star-forming {regions is used as a proxy for the dense gas fraction, $f_{dense}$, as in \citet[][]{2023MNRAS.521.3348N}}.

{ In the study of dense gas tracers the HCN luminosity ($L_{HCN}$)  {can be converted to dense gas mass ($M_{dense}$), which has been linked} by a tight correlation with the far-infrared luminosity, which is closely correlated to the star formation rate ($SFR$): $L_{HCN} \propto M_{dense} \propto L_{FIR} \propto SFR$ \citep[see for example,][]{2004ApJS..152...63G, 2019ApJ...880..127J, 2023MNRAS.521.3348N}. This  relationship has been observed in individual molecular clouds in the Milky Way and  nearby galaxies \citep{2005ApJ...635L.173W, 2006ApJ...640L.135G, 2008AJ....136.2846B, 2010ApJS..188..313W, 2015AJ....150..115U, 2016ApJ...822L..26B,2019ApJ...880..127J}, and also in  few  distant galaxies ($z\geq$1). Before \citet[][]{2022A&A...667A..70R} only two SF galaxies and three quasar hosts exhibited HCN emissions. These authors showed the difficulties of detecting  HCN at $z$=2.5-3.3, but concluded that the HCN/FIR ratios found are consistent with  normal starburst galaxies, not ultraluminous ones (c.f., their Figure~5).}

 {Additional information about the properties of the gas in the central kiloparsecs of galaxies can be obtained studying other emission mechanisms that occur inside those spatial scales, such as AGN.}{ According to the { Unified AGN model}  
\citep[see][]{1993ARA&A..31..473A, 1995PASP..107..803U}, 
the key to distinguishing between Seyfert 1 and 2 galaxies is the existence of a dust structure surrounding the central engine that obscures the inner parts of the AGN on the line of sight (LOS). Based on its geometric shape this structure is called a dusty torus; however, different distributions have been
 proposed for dust such as that proposed by  \citet[e.g.,][]{2006MNRAS.366..767F,articleNen, 2017ApJ...838L..20H}.
The contribution of the dusty torus luminosity to the integrated spectral energy distribution (SED) could be estimated by an SED fitting analysis \citep[e.g.,][]{2015A&A...576A..10C, Miyaji2019} which yields the dusty torus properties together with the galaxy continuum and line-emitting components, as well as the contribution of the AGN.} 
As part of a study of nearby galaxies with obscured AGN, we analyzed  the molecular gas spectra in the 3~mm band of the central, 1.6 kpc in diameter, region of the galaxy NGC\,4303 (M 61). NGC\,4303  is a barred spiral in the Virgo supercluster  at $z=$0.00522 ($D_L=$ 16.99 Mpc; 1$\arcsec$= 81 pc \citep{2006PASP..118.1711W}), with  an inclination of 25$^\circ$ \citep{1996AJ....111..174F}, classified as SAB(rs)bc  \citep{1991rc3..book.....D}. Its nuclear activity  has been debated, as \citet{1982AJ.....87.1628H} classified it as a LINER, whereas \citet{1985ApJS...57..503F} as a Seyfert 2,  which was confirmed by \citet{1986A&AS...66..335V} based on  an  equivalent width (EW) of [NII] of 400 km s$^{-1}$. \citet{2003ApJ...593..127J} used Chandra images with a central radius of 3 pc and showed that in this X-ray region, both a core source (either an AGN or ultraluminous x-ray source) with hydrogen column density $N_H\sim$1$\times$10$^{20}$~cm$^{-2}$, and an annular starburst region of radius 3$\arcsec~$ which is more obscured with $N_H\sim$5$\times$10$^{21}$~cm$^{-2}$, coexist. The AGN characterization of NGC\,4303 was confirmed using an infrared color-color diagram  \citep{2015A&A...578A..48C} and BPT diagrams \citep[][]{2017ApJ...846..102M}. Furthermore,  \citet{2020ApJ...905...29E} suggested that the nucleus is a candidate for the AGN fading phase. 

{ {The AGN contribution in NGC\,4303 will be studied by analyzing the  continuum emission} from UV to far-infrared wavelengths. The dusty torus properties provide important parameters that  complement the molecular gas characterization of the central 1.6 kpc region studied in this paper. Our aim is to obtain parameters of the interstellar medium in NGC\,4303 such as the molecular gas mass, HCN gas luminosity, far-infrared luminosity and star formation rate. 

The remainder of this paper is organized as follows: Section 2 describes the observational details of the 3 mm band spectroscopy. Section 3 presents the data reduction. An analysis of the molecular line emission spectra of NGC\,4303  is presented in \S 4. The  spectral energy distribution (SED) characterization to study  the dusty torus and AGN contribution is presented in \S 5. The discussion of results is presented in \S 6. A summary of the main conclusions and  final remarks is presented in \S 7. Throughout our work we adopt a cosmology where  $H_{0}$=69.6 ~km\,s$^{-1}$\,Mpc$^{-1}$, $\Omega_{m}$=0.286 and $\Omega_{\lambda}=$0.714 \citep{2014ApJ...794..135B}.}


\section{Observations}

Observations of NGC\,4303 were obtained on June 28 and December 22, 2015, with the Redshift Search Receiver (RSR) on the Large Millimeter Telescope Alfonso Serrano (LMT), hereon RSR/LMT, in its early science phase which operated with a 32-m active surface. Observing conditions at the  Volc\'an Sierra Negra site was good most of the time. During the observations the opacity at the  Volc\'an Sierra Negra site at a frequency of 225 GHz was in the range of $\tau$\,=\,0.18 - 0.19. The  system temperature, $T_{sys}$, was in the range of 85 and 110 K. Observations were centered on the coordinates of the galaxy nucleus ( { $\alpha_{2000}$: 12$^{h}$:21$^{m}$:54$^{s}$.9, $\delta_{2000}$: $+$04$^{o}$:28$^{\prime}$:25$^{\arcsec})$, with the OFF beam 39 arcsec apart. The pointing accuracy was  better than 2$\arcsec$. The total ON source integration time of NGC\,4303 was 1 hr.}

{The RSR is an autocorrelator spectrometer with a monolithic microwave integrated circuit system that receives signals over four pixels, simultaneously covering the frequency range 73 to 111 GHz at $\Delta\nu$\,=\,31\,MHz spectral resolution, which corresponds to $\sim$100\,\kms\ at 90 GHz \citep[see][]{2007ASPC..375...71E}.}
Hence, across the whole 3~mm band, the velocity resolution changes  {from 125\,\kms\, to 85 \,\kms}. {The RSR/LMT-32m has  {an angular} resolution or beam full width at half-maximum (FWHM) that is also frequency dependent, ranging from 28 to 19 arcseconds, between frequencies 73 and 111 GHz.}

\section{Reduced Spectra}\label{RSRspectra}

Autocorrelations, spectral co-adding, calibration, and baseline removal were performed using the \textit{Data \, REduction and Analysis \, Methods in Python} (DREAMPY) software developed by Gopal Narayanan for the RSR. After removing integrations with unstable bandpass, the remaining spectra were averaged{,} weighted by the $RMS$ noise in each individual spectrum. A simple linear baseline was removed in each spectral chassis, which covered a section of 6.5 GHz of the total RSR bandwidth after masking strong emission lines. To convert the antenna temperature ($T_A$) units to flux units (Jy), we used a conversion factor \textit{G}, the gain of the {LMT-32m}, given by  
\begin{equation}
    \label{GainRSR}
    G(\nu) = 7 \, \left( \dfrac{\nu \, [GHz]}{100 \, [GHz]}\right)\, [Jy\,K^{-1}].
\end{equation}

The spectral reduction last step was to obtain the main beam temperature ($T_{mb}$). To convert antenna temperature to $T_{mb}$, we divided the spectrum by the main beam efficiency ($\eta_{mb}$) 
\begin{equation}
    \label{EfimbGTM}
    \eta_{mb} = \left( 1.2 \, \exp(\nu/170)^2 \right) ^{-1}
\end{equation}
where $\nu$ denotes the sky frequency of the observed line. Such expression, Eq.~\eqref{EfimbGTM}, was obtained from calibration observations performed in June and December 2015. Details of the RSR/LMT data reduction can be found in several studies, for example \citet{2011AJ....141...38S,2014MNRAS.441.1363L,2015MNRAS.454.3485Y,2016MNRAS.459.3287C}. 

 {The RSR covers a frequency range of 84 - 111 GHz in the rest frame of NGC\,4303}. The entire RSR spectrum is shown in  Fig.~\ref{fig:lmtspectra}, together with the identified lines.  The {\it  RMS} of the spectrum varies slightly across the frequency band  from 0.618 mK at the low-frequency end 
 ($\nu>$ 92 GHz) to 0.854 mK at the high-frequency end. Several molecular spectral lines are clearly detected in this spectrum.  Using a 3$\sigma$ detection threshold on the line integrated intensity  and rest frequencies from the \textit{Splatalogue}\footnote{\url{https://splatalogue.online/#/home}}, we detected  { seven} spectral lines:
C$_2$H   at 87.31 GHz N=1--0, which is a blend of 6 lines: J=3/2-1/2 (F=1-1, 2-1, 1-0) and J=1/2-1/2 (F=1-1, 0-1, 1-0); HCN(1-0), HCO$^+$(1-0), HNC(1-0), CS(2-1), C$^{18}$O(1-0), and $^{13}$CO(1-0) at 88.63, 89.18, 90.66, 109.78 and 110.20GHz, respectively 
and the possible detection of three lines: N$_2$H$^+$ at 93.17 GHz, NH$_2$CN at 99.75 GHz and $^{13}$CN at 108.65 GHz.

 {We note that the OFF beam position of the RSR/LMT observations falls within the main body of the  galaxy; however, after careful inspection of the reduced data, we found that the $^{13}$CO(1-0) spectra} do not exhibit an absorption feature, which would appear if the 39 arcsec OFF beam was affected by the OFF galaxy emission. Contamination from the galaxy in the OFF position would be produced by faint CO emission in the spiral arms or interarm regions, c.f., PHANGS-ALMA CO(2-1) image of NGC\,4303 in \citet{2021ApJS..257...43L}
and would affect all lines, dense and diffuse. This is not found in any of the lines detected, so the contamination, if present, is  small.}

\begin{figure*}
\centering
\includegraphics[width=\textwidth]{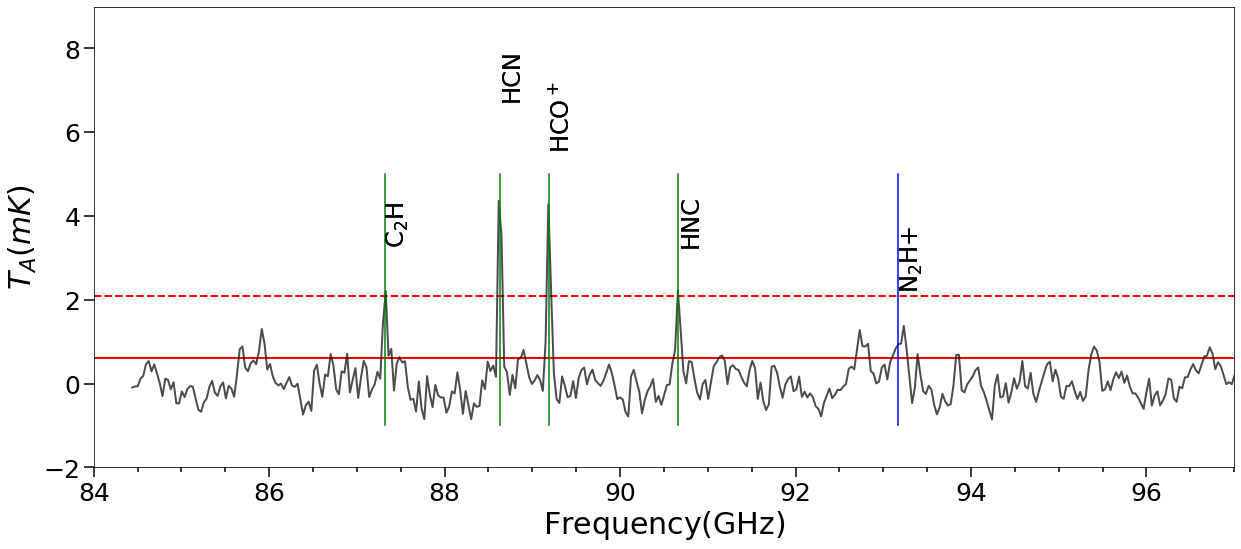} \\
\includegraphics[width=\textwidth]{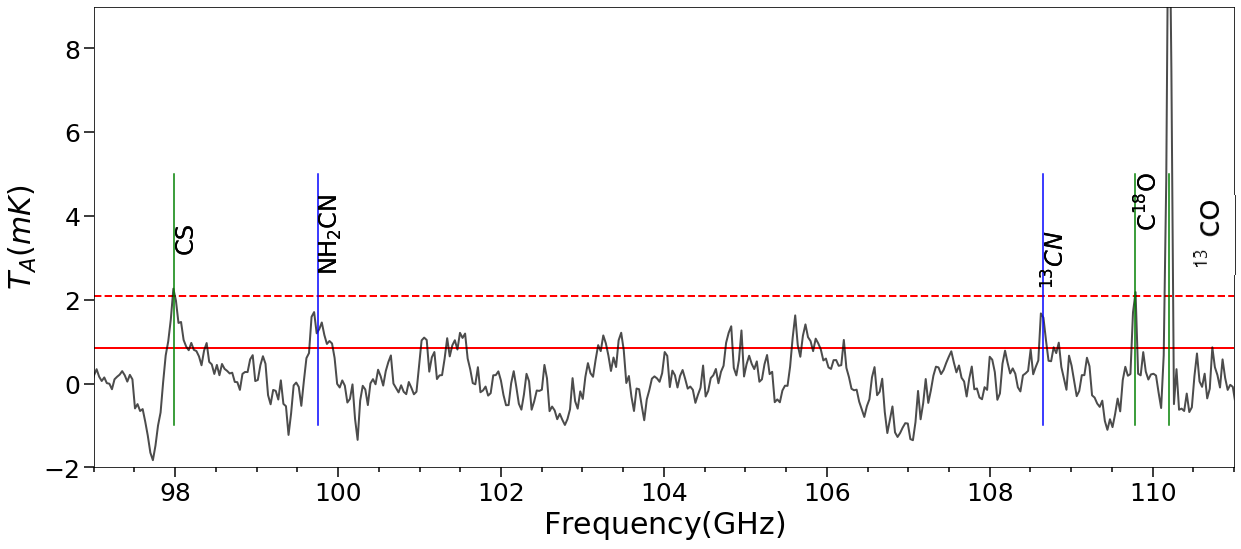} \\
    \caption{NGC\,4303  rest-frame spectra from 84 to  {111} GHz  obtained with the Redshift Search Receiver (RSR/LMT). The vertical color lines indicate the likely  frequencies of some molecular transitions.  {The red solid line indicates the  root mean square (RMS) value in each panel. The red dashed line represents the value of $3\sigma$.} Note that the detected transitions,  {vertical green lines, are detections of C$_2$H, HCN, HCO$^+$, HNC, CS(2-1), C$^{18}$O, and $^{13}$CO, whereas the possible detections in the vertical blue lines are N$_2$H$^+$, NH$_2$CN and $^{13}$CN} (see Table~\ref{table:molecules}).} \label{fig:lmtspectra}
\end{figure*}
\begin{figure*}
\centering
\includegraphics[width=\textwidth]{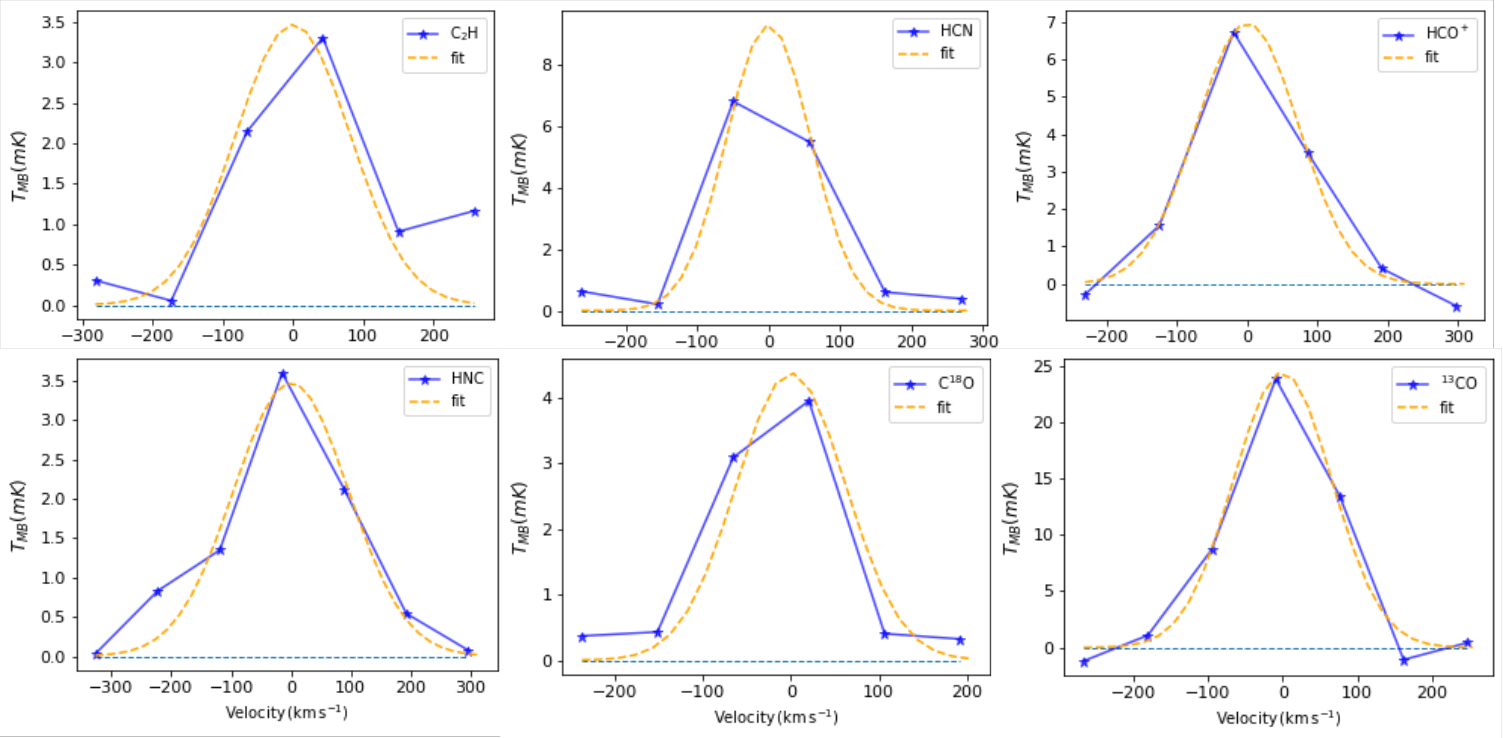}
\caption{Observed (blue) molecular lines with $S/N>3$ of NGC\,4303 and their Gaussian fits (orange dashed lines). We detected dense gas tracers as well as $^{12}$CO isotopic varieties. From left to right the molecular lines are as follows: (top row) C$^2$H, HCN, HCO$^+$; (bottom row) HNC, C$^{18}$O and $^{13}$CO.} 
\label{fig:indplots}
\end{figure*}


\section{Analysis}\label{analysis}

\subsection{Molecular gas}

 {Seven molecular lines were measured to have integrated line intensity with $S/N>3$, but CS(2-1) which shows a drop at the 
 low-frequency end, is considered a marginal detection}, and so is not considered in the line analysis. {For each line, a single Gaussian fit was performed to obtain the integrated intensity, {calculated in a range of $-250$ to $+250$ km s$^{-1}$, centered on their resting frequencies.} The error was computed using the covariance matrix of the fit. The FWHM and  peak intensities were obtained using fitting parameters. Individual plots of the detected lines are shown in Fig.~\ref{fig:indplots} and the derived parameters are listed in Table~\ref{table:molecules}}.

The most intense line is $^{13}$CO(1-0) followed by  HCN, HCO$^{+}$ and HNC.   {In a previous study on molecular gas in NGC\,4303, \citet{2019PASJ...71S..13Y} detected  {similar integrated intensities than those reported here for} $^{12}$CO(1-0) and $^{13}$CO(1-0) using the Nobeyama 45 m radio telescope,  {but they were not able to detect} C$^{18}$O(1-0). Also, \citet{10.1093/mnras/stab745} detected the same dense gas tracers lines (HCN, HCO$^+$, HNC) with the IRAM 30 m  {and reported similar integrated measurements to those  of this work}.}

 {In the following steps, we calculate the mass of the dense molecular gas in the central 1.6 kpc. First, with a combination of optically thin and thick lines, optical depth can be derived. {Then, we calculate the column density to finally use it in the calculation of the molecular gas mass}.} Following \citet{2017MNRAS.466...49J},{ assuming local thermodynamic equilibrium (LTE), solar abundances and a given excitation temperature in a cloud,} the optical depth ($\tau_{13}$) for $^{13}$CO can be calculated from the integrated intensities of $^{12}$CO and $^{13}$CO, $I_{12}$ and $I_{13}$, respectively. {Assuming the beam-filling factor of each line to be the same in the equation of radiative transfer, the optical depth is:} 
\begin{equation}
    \label{eq:tau13}
    \begin{split}
       \tau_{13} = -\ln \left[ 1-\dfrac{I_{13}}{I_{12}} \right]
    \end{split}
\end{equation}
{Because $^{12}$CO for NGC\,4303 cannot be detected in the frequency range of our observations}, we used the value $I_{12}$\,=\,55.2 $\pm$ 5.5 K\,km\,s$^{-1}$ 
obtained with the EMIR spectrograph at the IRAM 30m telescope  {(FWHM beam size 22${\arcsec}$}) reported by \citet{2020A&A...635A.131I} and  $I_{13}$ derived from our spectra\footnote{{ The difference in beam size of 3\arcsec\ between the two telescopes  is small, and might not strongly affect the LMT value of $I_{13}$ used, also it is unlikely that 
$\tau_{13}>$1.}}. By using the values in Eq.~\eqref{eq:tau13} yields $\tau_{13}$\,=\,{0.09}$\,\pm\,$0.01, that is, an optically thin line.  For $^{12}$CO{,}  {its optical depth is calculated by multiplying $\tau_{13}$ by} the relative  {solar} abundance\footnote{ { This is the value that most studies assume, but it can vary by a factor of 4.5 between galaxy centers, the local ISM or the solar system.}} [$^{12}$CO/$^{13}$CO]\,=\,89 \citep{1994ARA&A..32..191W},  which yields $\tau_{12}$\,=\,7.12 $\pm$ 0.89; {that is,  the $^{12}$CO line is} optically thick.

Now, we used the integrated intensity $I_{13}$ and $\tau_{13}$ to obtain the column density $N_{13}$. {Eq.~\eqref{eq:N13}} is the result of using the radiative transfer equation to calculate the number of molecules of $^{13}$CO over a path length \citep[see eq.1.7 in][]{2017PhDT.......113J}:
\begin{equation}
    \label{eq:N13}
    \begin{split}
        \left[ \dfrac{N_{13}}{cm^{-2}} \right] = 3\times 10^{14} &\left[ \dfrac{I_{13}}{K\, km\,s^{-1}} \right] \left[ \dfrac{1}{1-e^{-\tau_{13}}} \right] \\ &\dfrac{\tau_{13}}{1-e^{-5.29/T_{ex}}},
    \end{split}
\end{equation}
Three excitation temperature ($T_{ex}$) values  commonly found in dense molecular clouds, 10, 20 and 30 K, are assumed in {Eq.~\eqref{eq:N13}} for the subsequent calculations. 

To obtain the mass of $H_2$, it was necessary to convert  $N_{13}$ to $N(H_2)$ using the following equation: 
\begin{equation}
    \label{eq:NH2}
    \begin{split}
        \left[ \dfrac{N(H_2)}{cm^{-2}} \right] = \left[ \dfrac{H_2}{^{13}CO} \right] \left[ \dfrac{N_{13}}{cm^{-2}} \right].
    \end{split}
\end{equation}
where the relative abundance was [H$_2$/$^{13}$CO]\,=\,588,235 \citep{1978ApJS...37..407D}. Finally, the $H_2$ mass was obtained from {Eq.~\eqref{eq:masaNube}} assuming a spherical cloud with a diameter calculated for the telescope main beam, which for the $^{13}$CO line is 20$\arcsec$, and that at the NGC\,4303 distance corresponds to a diameter $d\,=\,$1.647 kpc and $m_{H_2}$ is the molecular mass of H$_2$.
\begin{equation}
    \label{eq:masaNube}
    \begin{split}
        M(H_2) = \dfrac{\pi d^2}{4} N(H_2)\,m_{H_2}
    \end{split}
\end{equation}
 The derived parameters for NGC\,4303 are presented in Table~\ref{NM}, which includes the column densities $N_{13}$, $N_{12}$, $N(H_2)$ and mass $M({H_2})$ for the three values of $T_{ex}$. 

\begin{table*}[!t]
 \caption{Molecular line transitions detected in NGC\,4303}
\label{table:molecules}
\begin{center}
  \begin{tabular}{ccccc}
\hline\hline
     Molecule  & Frequency$^b$ &  Integrated Flux & $\Delta v$$^c$ & Peak Intensity$^d$  \\
      &  (GHz) &  (K km s$^{-1}$) & (km s$^{-1}$) & (mK)\\
     (1) & (2) & (3) & (4) & (5)  \\\hline
C$_2$H(1-0)     & 87.31  & 0.77$\pm$0.03 & 192$\pm$38 &  3.47$\pm$0.96 \\
HCN(1-0)        & 88.63  & 1.55$\pm$0.11 & 138$\pm$59  & 9.28$\pm$0.97 \\
HCO$^+$(1-0)    & 89.18  & 1.41$\pm$0.09 & 171$\pm$73  & 6.98$\pm$0.97 \\
HNC(1-0)        & 90.66  & 0.90$\pm$0.04 & 225$\pm$48  & 3.47$\pm$0.98 \\
CS(2-1)$^a$     &  97.98 &  $<$2.03$\pm$0.68$^e$ & -- &  -- \\
C$^{18}$O(1-0)  & 109.78 &  0.77$\pm$0.04 & 146$\pm$28  & 4.37$\pm$1.54 \\
$^{13}$CO(1-0)  & 110.20 &  4.53$\pm$ 0.26 & 155$\pm$29 &  24.45$\pm$1.55  \\
\hline\hline
\end{tabular}
\begin{minipage}{0.95\textwidth}%
Notes:\\ (a) Possible detection. \\
(b) Rest frequencies, obtained from the Splatalogue database at \url{https://splatalogue.online/#/home}.\\
(c)  FWHM in velocity units. \\
(d) Peak intensity of main beam temperature (T$_{MB}$).\\
 {(e) Upper limit, because line has a blue absorption feature the integrated flux and error result large (see Fig.\ref{fig:lmtspectra}).}\\
  \end{minipage}%
\end{center}
\label{table1}
\end{table*}


\begin{table*}[!t]
 \caption{Column densities of  $^{12}$CO, $^{13}$CO and H$_2$  and $M({H_2})$}
\label{NM}
\begin{center}
  \begin{tabular}{ccccc}
\hline\hline
    $T_{ex}$ & $N_{13}$ & $N(H_2)$ & $M(H_2)^\ast$ \\ 
    $[$K$]$ & [$\times10^{15}$cm$^{-2}$]  & [$\times10^{21}$cm$^{-2}$] & [$\times10^{7}M_{\odot}$] \\
     \\ \hline
10 & 3.44 $\pm$ 0.63 &  2.02 $\pm$ 0.37 & 6.88 $\pm$ 1.26 \\
20 & 6.08 $\pm$ 1.11 & 3.58$\pm$ 0.65  & 12.17 $\pm$ 2.23 \\
30 & 8.74 $\pm$ 1.60 & 5.14 $\pm$ 0.94 & 17.50 $\pm$ 3.20 \\ 
\hline\hline
\end{tabular}
\begin{minipage}{0.59\textwidth}%
 {Note:\\ $^\ast$$M(H_2)$ was calculated from Eq.~\eqref{eq:masaNube}}.
  \end{minipage}%
\end{center}
\label{table2}
\end{table*}


To obtain the total molecular gas mass ($M_{mol}$), the contribution of molecular helium to molecular hydrogen must be included ~ in the relationship ~$M_{mol}\sim$1.36~ $M(H_2)$ from \citet{2022ApJ...925...72T}. Therefore, for $T_{ex}=$ 30 K  we obtain $M_{mol}$\,=\, (2.38\,$\pm$\,0.43)$\times10^{8}$ M$_{\odot}$.  {To derive the amount of gas in the observed region, we calculated the surface density of the molecular gas ($\Sigma_{mol}$), which is the mass of H$_2$ per 
square parsecs area. Taking $M_{mol}$ and the equivalent area of a circle of 1647 pc in diameter, the total surface density of  molecular gas is $\Sigma_{mol}$\,=\,112\,$\pm$\,20  M$_{\odot}$ pc$^{-2}$.}

 {Using the same procedure as described above, we obtained  the total molecular gas mass and  surface density of the molecular gas for $T_{ex}=$ 10 K, that is, $M_{mol}$\,=\, (9.35\,$\pm$\,1.71)$\times10^{7}$ M$_{\odot}$ and $\Sigma_{mol}$\,=\,46.5\,$\pm$\,8.5 M$_{\odot}$ pc$^{-2}$. Within this range of possible values of $T_{ex}$,  $\Sigma_{mol}$ varies by a factor of 2.}

\citet{2015AJ....150..115U} calculated the  surface density of molecular gas  from the intensity of $^{12}$CO using the following equation:
\begin{equation}
    \label{DenSmolecular}
        \Sigma_{mol} [M_{\odot}\, pc^{-2}]= \alpha_{CO} \,I_{12} \cos(i),
\end{equation}
where $i$ is the inclination of the galaxy, and $\alpha_{CO}$ is the conversion factor of $^{12}$CO to molecular {  gas mass with an adopted value  of} $4.4$ M$_{\odot} $pc$^{-2}$(K km s$^{-1})^{-1}$ {  (including the 1.36 factor for helium) with a $\pm$30\% uncertainty} given by \citet{2013ARA&A..51..207B}. Using $I_{12}$ from literature, we obtain  {$\Sigma_{mol}= 220\pm 79 $} M$_{\odot}$ pc$^{-2}$.

Analogously, we calculated the surface density of the dense molecular gas traced by the HCN line using the equation:
\begin{equation}
    \label{DenSdenso}
        \Sigma_{dense} [M_{\odot}\, pc^{-2}] = \alpha_{HCN}\, I_{HCN} \cos(i)
\end{equation}
where $\alpha_{HCN}= 10$ M$_{\odot}$ pc$^{-2}$ (K km s$^{-1})^{-1}$, proposed by \citet{2004ApJ...606..271G}  {is} a typical {upper limit}  for spiral galaxies\footnote{ {\citet{2017ApJ...835..217L} reported an extensive analysis of the use of millimeter-wave emission line ratios to trace molecular gas density. In the dense gas conversion factors, they considered a range of densities for the models as well as the distribution of widths and temperatures}}. The result is $\Sigma_{dense}$=14$\pm$1 M$_{\odot}$ pc$^{-2}$.

 {We calculate the dense gas mass as traced by HCN, $M_{dense}$, by  multiplying} $\Sigma_{dense}$ by the area of the telescope's main beam, with a diameter of 25$\arcsec$  for the HCN line. The result is: $M_{dense}=(4.7\pm0.3) \times 10^{7}$ M$_{\odot}$.

Finally, with $\Sigma_{dense}$ and $\Sigma_{mol}$ for $T_{ex}=30$ K, we can calculate the  {dense molecular fraction} ($f_{dense}$) in the observed region as:
\begin{equation}
    \label{Fdenso}
        f_{dense}= \dfrac{\Sigma_{dense}}{\Sigma_{mol}}=0.13\pm0.06
\end{equation}

\subsection{Luminosities}

\citet{2004ApJ...606..271G} calculated the luminosity of an extended source with a size larger than the main beam of the telescope. Because  the galaxy is not mapped on the major axis, it is considered  a source with a size equal to or smaller than the telescope's main beam. The calculation is given by:
\begin{equation}
\begin{split}
    \label{eqn:lumHCN}
    L'_{HCN}\,[K \, km\,s^{-1}\,pc^{2}] &\approx \Omega_{mb} \,I_{HCN} \,D_L^2 \,(1+z)^{-3} 
\end{split}
\end{equation}
where $\Omega_{mb}$ is the solid angle of the main beam, $D_L$(pc) is the luminosity distance, $I_{HCN}$ is the integrated HCN line  intensity  and $z$ is the \textit{redshift}. We can rewrite $\Omega_{mb}$ in terms of the antenna parameters, that is, the FWHM of the main beam $\theta_{\mathrm{FHWM}}$ (in radians) and obtain the HCN luminosity as
\begin{equation}
\begin{split}
    \label{LUM}
        L'_{HCN}\,[K \, km\,s^{-1}\,pc^{2}] \approx \dfrac{\pi}{4 \, ln(2)}\theta^2_{\mathrm{FHWM}} \,I_{HCN} \\\,D_L^2 \, (1+z)^{-3}.
\end{split}
\end{equation}

We then calculate the luminosity of the dense gas tracers substituting the integrated intensities of this work, e.g., HCN by HCO$^+$ and the corresponding values of $\theta_{\mathrm{FHWM}}$  {at the observing frequency, $\sim$25$\arcsec$ for the dense gas tracers and 20$\arcsec$ for $^{13}$CO  (see Table~\ref{tab:lum}).}

If we use the mass-luminosity conversion factor defined as $\alpha_{_{13}CO}$ with units M$_{\odot}$ pc$^{-2}$ (K km s$^{-1})^{-1}$, one can rewrite the Solomon relation  {\citep{1997ApJ...478..144S}} as $M(H_2)= \alpha_{_{13}CO}\, L^{\prime}_{13}$ or
\begin{equation}
    \label{alfa13}
    \alpha_{_{13}CO} = \dfrac{\Sigma_{H_2}}{I_{13}}.
\end{equation}

Using the  previously calculated data for $M(H_{2})$, $\Sigma_{H_2}=$82\,$\pm$\,15  M$_{\odot}$ pc$^{-2}$ and  Eq.~(\ref{alfa13}), we calculated both the factor $\alpha_{_{13}CO}$ and the $^{13}$CO luminosity, resulting $\alpha_{_{13}CO} = (18.2\pm 4.4)$ M$_{\odot}$ pc$^{-2}$ (K km s$^{-1})^{-1}$ and $L^{\prime}_{13}= (9.6\pm 3.9 )\times10^{6}$ K km s$^{-1}$ pc$^{2}$. The $^{12}$CO luminosity is calculated with the Eq.~(\ref{DenSmolecular}) and the FWHM of 22$\arcsec$ of the observed line, which  results in $L^{\prime}_{12}= 1.29\times10^{8}$ K km s$^{-1}$ pc$^{2}$.

{To estimate the abundance of molecular gas in the nuclear obscured region, we used the integrated intensity of the different  emission lines of  dense gas tracers ({C$_2$H}, HCN, HNC and HCO$^+$),  diffuse gas tracers ($^{13}$CO, C$^{18}$O) and   $^{12}$CO integrated intensity  $I_{12}=55.2 \pm 5.5$ K km s$^{-1}$  \citep{2020A&A...635A.131I} {to calculate their line ratios} (Table~\ref{tab:data2}}).


\begin{table*}[!t]
 \caption{Luminosities for diffuse and dense gas emission lines in NGC\,4303.}
\label{tab:lum}
\begin{center}
  \begin{tabular}{lcc}
\hline\hline
		Line  & L$_{gas}$    & $\log(L_{gas}$) \\
                  &  $\times10^{6}\,\,[K\,\,km\,\,s^{-1}\,pc^{2}]$          &   $[K\,\,km\,\,s^{-1}\,pc^{2}]$       \\
		\hline
               $^{13}$CO & 9.60 $\pm$ 3.90 & 6.98 $\pm$ 0.09\\ \hline
             C$_2$H & 3.78 $\pm$ 0.68 & 6.58 $\pm$ 0.07\\
             HCN & 7.38 $\pm$ 1.40 & 6.87 $\pm$ 0.08\\
             HCO$^{+}$ & 6.63 $\pm$ 1.24 & 6.82 $\pm$ 0.08\\
             HNC & 4.09 $\pm$ 0.75 & 6.61 $\pm$ 0.08\\ \hline
             Total dense & 21.88 $\pm$ 4.07 & 7.34 $\pm$ 0.08\\ 
\hline\hline
\end{tabular}
\end{center}
\end{table*}

\begin{table}[!t]
 \caption{Abundances of dense gas relative to \\diffuse  and dense gas tracers.}
 \label{tab:data2}
\begin{center}
  \begin{tabular}{lc}
\hline\hline
Ratio$^\ast$      & Value \\
                  &                      \\
		\hline
            C$_2$H / C$^{18}$O& 1.01  $\pm$ 0.07 \\
HCN/ C$^{18}$O  &   2.01   $\pm$ 0.18\\
HCO$^{+}$/ C$^{18}$O & 1.83  $\pm$ 0.15  \\
HNC / C$^{18}$O&   1.17 $\pm$ 0.08 \\
CS / C$^{18}$O & 2.67   $\pm$ 0.19\\
$^{13}$CO/ C$^{18}$O & 5.88 $\pm$ 0.46 \\
HCN / $^{12}$CO   &  0.028 $\pm$ 0.002 \\
HCO$^+$ / $^{12}$CO   &  0.025 $\pm$ 0.003  \\
HNC / $^{12}$CO   &  0.016 $\pm$ 0.003 \\
$^{13}$CO / $^{12}$CO   &  0.082 $\pm$ 0.010 \\
C$^{18}$O / $^{12}$CO   &  0.013 $\pm$ 0.002\\
        \hline\hline
        HNC/ HCN  &   0.58 $\pm$ 0.04   \\
HCO$^+$ / HCN & 0.91  $\pm$ 0.08   \\
C$_2$H / HCN &   0.38  $\pm$ 0.04 \\
\hline\hline
\end{tabular}
\begin{minipage}{0.95\columnwidth}%
\begin{center}
Note: $^\ast$ For $^{12}$CO we used \\
$I_{12}$\,=\,55.2$\pm$5.5 K km s$^{-1}$  \citep{2020A&A...635A.131I}.\end{center}
\end{minipage}
\end{center}
\end{table}

\subsection{Dusty torus, SFR and AGN contribution to the SED}

As mentioned in \S~\ref{sec:intro},  {previous} multiwavelength 
studies  {confirmed that} the nuclear activity  {of NGC\,4303 is} influenced by {an accreting} SMBH \citep{1985ApJS...57..503F,1986A&AS...66..335V,2017ApJ...846..102M,2015A&A...578A..48C}, classifying it as Sy 2 \citep{2006A&A...455..773V}. 
{In this work we are also interested in the physical properties of NGC\,4303, mainly the star formation rate (SFR), {in order to compare the values derived here with those from 
the literature.} A good tool to achieve this task is  integrated spectral energy 
distribution (SED) fitting analysis, which includes an AGN component. Interestingly, { this }{complementary} analysis allows us to obtain not only the galactic 
properties but also the AGN properties, including dusty torus characteristics. All 
{these} properties are useful for follow-up studies.}

For this purpose, we used the Code Investigating GALaxy Evolution{ , called CIGALE, developed by }\citet[][]{Boquien2019}, 
which allows fitting of the SED of galaxies using different models for the relevant physical 
components. For instance, it is possible to select from different star-forming history models, stellar populations, initial mass functions, attenuation laws, dust emissions and AGN models. The different components are fitted to preserve the energy balance principle, which means that the energy absorbed at UV and optical wavelengths 
is re-emitted at infrared wavelengths. 
To select the best fit, CIGALE performs a minimization of the $\chi^2$ statistics and
performs Bayesian analysis to determine the probability distribution of the 
physical parameters obtained from the fits.

We searched for photometric data of NGC\,4303 in the literature, using  {the VizieR 
photometry tool\footnote{http://vizier.cds.unistra.fr/vizier/sed/doc/}, which allows 
visualization of photometric points extracted from catalogues in ViZieR
around a sky position. We used a default} search radius of 5\arcsec~ {centered 
at NGC\,4303, which is small enough to consider data truly corresponding}
to the emission  {of the galaxy}, finding multi-wavelength observations 
at UV, optical, infrared and sub-millimeter wavelengths (see Table~\ref{tab:cigale_photometry}). 
{According to \citet{1991rc3..book.....D}, the isophotal dimensions ($R_{25}$) of 
NGC\,4303 are 6$\overset{\arcmin}{.}$46$\pm$0$\overset{\arcmin}{.}$15 and 5$\overset{\arcmin}{.}$75$\pm$0$\overset{\arcmin}{.}$27, for the major and minor axes, respectively. This
means that all the observations retrieved form the literature include the total emission of 
the galaxy.}

The UV data consist of  Galaxy Evolution Explorer \citep[GALEX;][]{Martin2005} observations 
in the far-ultraviolet (FUV; $\lambda_{\rm eff}$\,=\,0.153 $\mu$m) and  near-ultraviolet (NUV; 
$\lambda_{\rm eff}$\,=\,0.231 $\mu$m) bands  {with an angular resolution of 5$\overset{\arcsec}{.}$3
and  1$\overset{\circ}{.}$2 field of view (FOV)}, taken from \citet{2012yCat..35440101C}. The optical 
data consist of observations from the Sloan Digital Sky Survey Data Release 7 
\citep[SDSS-DR7;][]{Abazajian2009} in the five bands \textit{u, g, r, i} and \textit{z},
taken from \citet{2014ApJS..215...22K}.  {These data have an angular resolution of 
$\leq 1\overset{\arcsec}{.}$5, and  3$^{\circ}$ FOV}. 
In the IR regime we compiled data from different sources: 
the Two Micron All Sky Survey \citep[2MASS;][]{2006AJ....131.1163S}, the \textit{Spitzer}
Space Telescope \citep{werner2004}, the Infrared Astronomical Satellite 
\citep[IRAS;][]{1984ApJ...278L...1N}, the Wide-field Infrared Survey Explorer \citep[WISE;][]{wright10}, 
and the \textit{Herschel} Space Telescope \citep{Pilbratt2010}. 
In the case of 2MASS data, we use photometry in the J (1.2 $\mu$m), H (1.6 $\mu$m) and K (2.2 $\mu$m)
bands from \citet{2015yCat..22200006B}, \citet{2016yCat.7275....0D} and \citet{2019yCat.4038....0S}, 
respectively,  {all of them with an angular resolution of 2$\arcsec$ and a 8$\overset{\circ}{.}$5 FOV}. 
In the case of Spitzer data, we considered observations obtained with the InfraRed Array Camera 
\citep[IRAC;][]{fazio04} in the IRAC1 (3.6 $\mu$m), and IRAC2 (4.5 $\mu$m) bands from 
\citet{2010PASP..122.1397S}, and  IRAC4 (8 $\mu$m) band from \citet{2014yCat..35650128C}. 
 {These data have resolution of 1$\overset{\arcsec}{.}$7, 1$\overset{\arcsec}{.}$6, 
1$\overset{\arcsec}{.}$9, respectively, with a 5$\arcmin$ FOV for IRAC1 and IRAC2 bands, and  {5$\overset{\arcmin}{.}$2$\times$5$\overset{\arcmin}{.}$2} in the case of IRAC4.}
In the case of IRAS data, we use the photometry in the 12 $\mu$m band from 
\citet{2006ApJS..164...81M}, the 25 and 60 $\mu$m bands from \citet{2006ApJ...652.1068R}, and 
the 100 $\mu$m band from \citet{2001ApJ...554..803Y},  {with a resolution of 0$\overset{\arcmin}{.}$5, 
0$\overset{\arcmin}{.}$5, 1$\overset{\arcmin}{.}$0 and 2$\overset{\arcmin}{.}$0, respectively, and a common 63$\overset{\arcmin}{.}$6 FOV}. 
We also consider data in the MIPS 24 and 160 $\mu$m bands from \citet{2012yCat..74230197B}, 
 {whose angular resolution is 6$\arcsec$ with a 5$\overset{\arcmin}{.}$4 $\times$ 5$\overset{\arcmin}{.}$4 FOV}.
From WISE, we use observations in the W1 (3.4 $\mu$m) and W2 (4.6 $\mu$m) bands from 
\citet{2019yCat.4038....0S}, the W3 (12 $\mu$m) band from \citet{2014yCat..35650128C}, and 
the W4 (22 $\mu$m) band from \citet{2014yCat..35700069B}.  {These data have angular 
resolution of 6$\overset{\arcsec}{.}$1, 6$\overset{\arcsec}{.}$4, 6$\overset{\arcsec}{.}$5, 12$\overset{\arcsec}{.}$0, respectively, and a 
common 47$\arcmin$ FOV.}
In the case of \textit{Herschel} data, we use observations obtained with the Photodetector Array 
Camera and Spectrometer \citep[PACS;][]{2010A&A...518L...2P}, in the 100 $\mu$m (PACSgreen) and 
160 $\mu$m (PACSred) bands  {with angular resolution of 8$\arcsec$ and 13$\arcsec$, 
respectively, and a 3$\overset{\arcmin}{.}$5$\times$1$\overset{\arcmin}{.}$75 FOV, as well as} the Spectral and Photometric 
Imaging Receiver \citep[SPIRE;][]{Griffin2010}, in the 250 $\mu$m (PSW), 350 $\mu$m (PMW) and 
500 $\mu$m (PLW) bands  {with angular resolution of 18$\arcsec$, 25$\arcsec$ and 36$\arcsec$, 
with a 8$\overset{\arcmin}{.}$0$\times$4$\overset{\arcmin}{.}$0 FOV}. 
PACS data were collected from \citet{2013yCat..74281880A}, 
whereas the SPIRE data were collected 
from \citet{2018yCat..51550017C}. 
Finally, we used observations taken with the Submillimeter Common-User Bolometer Array 
\citep[SCUBA;][]{1999MNRAS.303..659H} instrument on the James Clerk Maxwell Telescope (JCMT). 
These observations in the  850 $\mu$m band were taken from \citet{2008ApJS..175..277D} 
 {with angular resolution of 22$\overset{\arcsec}{.}$9 and a 2$\overset{\arcmin}{.}$3 FOV}. 

We fitted the SED of NGC\,4303 using {typical choices of the components to fit for galaxies harbouring an AGN \citep[e.g.][]{Wang2020, 2021MNRAS.507.3070S},} a delayed star formation history with an optional exponential burst and stellar populations model by \citet{BruzualCharlot2003} using the \texttt{sfhdelayed} 
module with a \citet{Chabrier2003} initial mass function (\texttt{bc03}). We considered nebular 
emission  to account for any Lyman continuum emission 
(\texttt{nebular}). The attenuation 
law (\texttt{dustatt\_modified\_starburst}) is that of \citet{Calzetti2000}, whereas the dust emission 
is modeled using  \citet{Jones2017} models (\texttt{themis}). For the AGN component we use the
CLUMPY\footnote{\url{https://clumpy.org/}} models through the \texttt{nenkova2008} module, which 
considers the emission from a clumpy torus \citep{Nenkova2008a,Nenkova2008b}. This latter module 
has recently been used successfully in other SED fitting studies of AGNs given its flexibility 
regarding the geometrical parameters of the torus \citep{Miyaji2019, Yamada2023}. Each of the 
aforementioned modules, corresponding to the physical components, has its own parameters that can 
take several values  to form the grid of models to be fit to the SED. The physical components, corresponding modules, parameters, and values adopted in this analysis are listed in Table~\ref{tab:params_table}.

\begin{figure*}
\centering
\includegraphics[width=11.5cm]{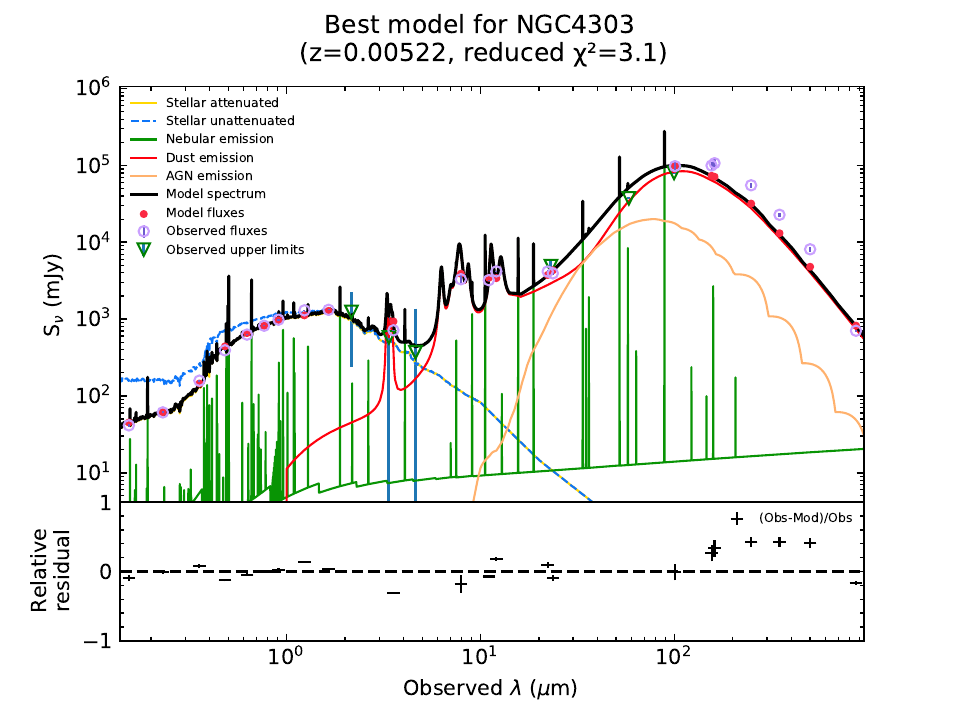}
\caption{The best SED model fit for NGC\,4303 using CIGALE. 
         { Shown are the} observed fluxes
         (purple circles),  model fluxes (red dots),  total model (black line),
         unattenuated stellar emission (blue line),  attenuated stellar contribution (yellow line),
          nebular emission (green line),  dust emission (red line) and  AGN emission (orange
         line). The bottom panel shows the residuals of the observed to the model fluxes.}
\label{fig:sed_fit}
\end{figure*}

In Figure~\ref{fig:sed_fit} we present the best fit for the NGC\,4303 SED. The observed 
fluxes (purple circles),  model fluxes (red circles), and  total model (black line) are shown. The different physical components used were as follows: the stellar component before (yellow line) and 
after attenuation (blue dashed line),  dust emission (red line),  nebular emission (green lines),
and  AGN component (appricot line). The flux residuals  in each band are shown in the bottom
panel and are calculated as $(S_{\rm \nu, obs} - S_{\rm \nu, model})/S_{\rm \nu, obs}$. The reduced
$\chi^2$ of the fit was 3.1.

Regarding the geometrical parameters of the clumpy dusty torus obtained with the best fit and 
the Bayesian analysis values (in parenthesis), we obtained a thick torus given its aperture above the equatorial plane 
\texttt{op\_angle} = 60$^\circ$ (56$^\circ \pm$9$^\circ$), a large optical depth 
\texttt{tau\_V}=200 (184$\pm$38), a large number of clouds
along the equatorial plane \texttt{N\_0}=15 (14$\pm$2), and an inner to outer radius ratio 
of \texttt{Y\_ratio}=100 (97$\pm$12). The power of the radial distribution of clouds 
($\propto r^{-q}$) is \texttt{q}=0.0 (0.12$\pm$0.29) and the viewing angle is \texttt{incl}=80$^{\circ}$ 
(67$^{\circ}\pm$16$^{\circ}$). By considering the classical unified AGN model 
\citep{1993ARA&A..31..473A,1995PASP..107..803U}. This viewing angle is in agreement with the {Sy 2}
classification of \citet{2006A&A...455..773V}.

The total AGN and torus luminosities of the best fit and the Bayesian analysis values (in parenthesis) are $L_{\rm AGN}$ = 1.5$\times$10$^{44}$ erg s$^{-1}$
((1.5 $\pm$ 0.6)$\times 10^{44}$ erg s$^{-1}$), 
$L_{\rm TORUS}$ = 6.975$\times$10$^{43}$
erg s$^{-1}$ ((7.1 $\pm$ 2.8) $\times 10^{43}$ erg s$^{-1}$), while the galaxy dust 
luminosity is $L_{\rm DUST}$ = 2.8$\times$10$^{44}$ erg s$^{-1}$ ((2.8 $\pm$ 0.1) $\times 10^{44}$ erg s$^{-1}$){, and the total IR luminosity is 3.5$\times10^{44}$ erg s$^{-1}$ (3.51 $ \pm $ 0.30)$\times10^{44}$ erg s$^{-1}$.} Another important quantity derived from CIGALE is the star formation rate, which in this case
is $SFR=$ 6.014 M$_\sun$ yr$^{-1}$ (5.969$\pm$ 0.298 M$_\sun$ yr$^{-1}$). {The parameters derived from CIGALE are presented in Table~\ref{tab:agn_results}}.

\begin{table*}[ht]
	\centering
	\caption{Photometry used in the SED fitting with CIGALE.}
    \begin{threeparttable}
	\label{tab:cigale_photometry}
	\begin{tabular}{lcccccc} 
		\hline
		Band      & $\lambda_{\rm eff}$ &  {Flux$^\dagger$} &Telescope/Instrument & Res. {$^{\dagger\dagger}$}  & FOV      & Ref. \\
                  & ($\mu$m)            &  {(mJy)} &or Survey           & ($\arcsec$) & ($\arcmin$) &   \\
		\hline
        FUV       & 0.153 & 41.3$\pm$1.9 & GALEX          & 5.3  & 72.0 & 1 \\
        NUV       & 0.231 & 61.0$\pm$1.7 & GALEX          & 5.3  & 72.0 & 1 \\
        u         & 0.352 & 158.0$\pm$5.0 & SDSS           & $\leq$ 1.5 & 180.0 & 2 \\
        g         & 0.482 & 392.0$\pm$7.0 & SDSS           & $\leq$ 1.5 & 180.0 & 2 \\
        r         & 0.625 & 628.0$\pm$12.0 & SDSS           & $\leq$ 1.5 & 180.0 & 2 \\
        i         & 0.763 & 819.0$\pm$15.0 & SDSS           & $\leq$ 1.5 & 180.0 & 2 \\
        z         & 0.902 & 982.0$\pm$28.0 & SDSS           & $\leq$ 1.5 & 180.0 & 2 \\
        J         &  1.2  & 1300.0$\pm$20.0 & 2MASS          & 2.0  & 8.5 & 3 \\
        H         &  1.6  & 1320.0$\pm$20.0 & 2MASS          & 2.0  & 8.5 & 4 \\
        Ks        &  2.2  & 1240.0$\pm$-999.0 & 2MASS          & 2.0  & 8.5 & 5 \\
        W1        &  3.35 & 578.0$\pm$-999.0 & WISE           & 6.1  & 47.0 & 5 \\
        IRAC1     &  3.6  & 714.0$\pm$1.0 & Spitzer/IRAC   & 1.7  & 5.0 & 6  \\
        IRAC2     &  4.5  & 479.0$\pm$1.0 & Spitzer/IRAC   & 1.6  & 5.0 & 6 \\
        W2        &  4.60 & 370.0$\pm$-999.0 & WISE           & 6.4  & 47.0 & 5 \\
        IRAC4     &  8.0  & 3300.0$\pm$410.0 & Spitzer/IRAC   & 1.9  & 5.2$\times$5.2 & 7 \\
        W3        &  12   & 4180.0$\pm$110.0 & WISE           & 6.5  & 47.0 & 7 \\
        IRAS1     &  12   & 3210.0$\pm$50.0 & IRAS           & 30.0 & 63.6 & 8 \\
        W4        &  22   & 4160.0$\pm$190.0 & WISE           & 12.0 & 47.0 & 9 \\
        MIPS24    &  24   & 3940.0$\pm$160.0 & Spitzer/MIPS   & 6.0  & 5.4$\times$5.4 & 10 \\
        IRAS2     &  25   & 4900.0$\pm$-999.0 & IRAS           & 30.0 & 63.6 & 11 \\
        IRAS3     &  60   & 37300.0$\pm$-999.0 & IRAS           & 60.0 & 63.6 & 11 \\
        PACSgreen &  100  & 97900.0$\pm$11800.0 & Herschel/PACS  & 8.0  & 3.5$\times$1.75 & 12 \\
        IRAS4     &  100  & 79700.0$\pm$-999.0 & IRAS           & 120.0 & 63.6 & 13 \\
        MIPS160   &  160  & 99800.0$\pm$12000.0 & Spitzer/MIPS   & 38.0 & 2.1$\times$5.3 & 10 \\
        PACSred   &  160  & 107000.0$\pm$13000.0 & Herschel/PACS  & 13.0 & 3.5$\times$1.75 & 12 \\
        PWS       &  250  & 55200.0$\pm$3900.0 & Herschel/SPIRE & 18.0 & 8.0$\times$4.0 & 14 \\
        PWM       &  350  & 22800.0$\pm$1600.0 & Herschel/SPIRE & 25.0 & 8.0$\times$4.0 & 14 \\
        PWL       &  500  & 8100.0$\pm$580.0 & Herschel/SPIRE & 36.0 & 8.0$\times$4.0 & 14 \\
        SCUBA2    &  850  & 705.0$\pm$20.0 & JCMT/SCUBA     & 22.9 & 2.3 & 15 \\
        \hline
	\end{tabular}
    \begin{tablenotes}
    \item[] REFERENCES: (1) \citet{2012yCat..35440101C}; (2) \citet{2014ApJS..215...22K}; (3) \citet{2015yCat..22200006B}; (4) \citet{2016yCat.7275....0D}; (5) \citet{2019yCat.4038....0S}: (6) \citet{2010PASP..122.1397S}; (7) \citet{2014yCat..35650128C}; (8) \citet{2006ApJS..164...81M}; (9) \citet{2014yCat..35700069B}; (10) \citet{2012yCat..74230197B}; (11) \citet{2006ApJ...652.1068R}; (12) \citet{2013yCat..74281880A}; (13) \citet{2001ApJ...554..803Y}; (14) 
    \citet{2018yCat..51550017C}; (15) \citet{2008ApJS..175..277D}. \\
    {$^\dagger$The flux data without error (-999.0) are taken as upper limits in the SED fitting analysis (green upside down triangles in Fig.~\ref{fig:sed_fit}).\\
    {$^{\dagger\dagger}$}Telescope resolution.}
    \end{tablenotes}
    \end{threeparttable}
\end{table*}
\begin{table*}[ht]
	\centering
	\caption{Parameters used in the SED fitting with CIGALE.}
    \begin{threeparttable}
	\label{tab:params_table}
	\begin{tabular}{llll} 
		\hline
		Component & Module & Parameter & Values \\
		\hline
	SFH	& \texttt{sfhdelayed} & \texttt{tau\_main}  & 1000, 2000, 4000, 6000 \\ & & & (Myr) \\
	      &                     & \texttt{age\_main}  & 1500, 4000, 8000 (Myr) \\	
        &                     & \texttt{tau\_burst} & 10, 25 (Myr)   \\
		  &                     & \texttt{age\_burst} & 10, 20 (Myr) \\
        &                     & \texttt{f\_burst}   & 0.0, 0.01 \\
		\hline
	Stellar	& \texttt{bc03}    & \texttt{imf}             &  \citet{Chabrier2003} \\
		emission  &                                  & \texttt{metallicity}     & 0.0001, 0.0004, 0.004,  \\ & & & 0.008, 0.02, 0.05 \\
		&                                  & \texttt{separation\_age} & 10 (Myr) \\
		\hline
	Nebular 	& \texttt{nebular} & \texttt{logU}            & -2.0 \\
		 emission &                                  & \texttt{f\_esc}          & 0.0\\
		  &                                  & \texttt{f\_dust}         & 0.0\\
		  &                                  & \texttt{lines\_width}    & 300.0\\
		\hline
	Attenuation 	& \texttt{dustatt\_modified}  & \texttt{E\_BV\_lines}    & 0.0, 0.3, 0.6, 0.9, 1.2,  \\ & & & 1.5, 1.8, 2.1, 2.4 \\
		law  & \texttt{\_starburst}                    & \texttt{E\_BV\_factor}              & 0.44 \\
		  &                    & \texttt{uv\_bump\_wavelength}       & 217.5 \\
		  &                    & \texttt{uv\_bump\_width}            & 35.0 \\
        &                    & \texttt{uv\_bump\_amplitude}        & 0, 1.5, 3 \\
		  &                    & \texttt{powerlaw\_slope}            & -0.2, 0 \\
		  &                    & \texttt{Ext\_law\_emission\_lines}  & 1 (MW) \\
        &                    & \texttt{Rv}  & 3.1 \\
		\hline
	Dust 	& \texttt{themis}     & \texttt{qhac}  & 0.06, 0.17, 0.36 \\  
		emission&                    & \texttt{umin}  & 0.1, 1.0, 10.0, 50.0  \\
		  &                    & \texttt{alpha} & 1.0, 2.0, 3.0 \\
		&                    & \texttt{gamma} & 0.1, 0.25 \\
		\hline
	AGN	& \texttt{nenkova2008}  & \texttt{Y\_ratio}     & 5, 10, 30, 60, 100 \\
		&                    & \texttt{op\_angle}     & 20, 40, 60 \\
	    &                    & \texttt{tau\_V}        & 20, 60, 80, 120, 200 \\
		&                    & \texttt{N\_0}          & 3, 5, 10, 15\\
		&                    & \texttt{incl}          & 10, 30, 60, 80\\
		&                    & \texttt{q}             & 0.0, 0.5, 1.0, 2.0, 3.0 \\
		&                    & \texttt{fracAGN} &  0.0, 0.1, 0.2, 0.3, 0.4,   \\ & & & 0.5, 0.6, 0.7, 0.8, 0.9 \\
		\hline 
 \end{tabular}
    \end{threeparttable}
\end{table*}

\begin{table*}[t]
	\centering
	\caption{Results for the AGN and dusty torus properties  \\after the SED fitting with CIGALE.}
	\label{tab:agn_results}
	\begin{tabular}{lcc} 
		\hline
		Parameter          & Best fit result & Bayessian analysis\\
        \hline
        \texttt{Y\_ratio}              &  100   &   97  $\pm$ 12 \\
		\texttt{op\_angle}($^{\circ}$) &   60   &   56  $\pm$ 9 \\
	    \texttt{tau\_V}                &  200   &  184  $\pm$ 38 \\
		\texttt{N\_0}                  &   15   &   14  $\pm$ 2  \\
		\texttt{incl}($^{\circ}$)      &   80   &   67  $\pm$ 16\\
		\texttt{q}                     &  0.0   &    0.12  $\pm$  0.29 \\
		\texttt{fracAGN}               &  0.2   &    0.20  $\pm$  0.07 \\
        $SFR$ (M$_\sun$ yr$^{-1}$)              &  6.0 &    6.0 $\pm$  0.3  \\
    $L_{\rm AGN}$ (erg s$^{-1}$)   & 1.5$\times10^{44}$ & (1.5 $ \pm $ 0.6)$\times10^{44}$ \\
    $L_{\rm TORUS}$ (erg s$^{-1}$) & 7.0$\times10^{43}$ & (7.1 $\pm $~2.8)$\times10^{43}$ \\
    $L_{\rm DUST}$ (erg s$^{-1}$)  & 2.8$\times10^{44}$ & (2.8 $ \pm $ 0.1)$\times10^{44}$ \\
     $L_{IR}= $ (erg s$^{-1}$)  & 3.5$\times10^{44}$ & (3.51 $ \pm $ 0.30)$\times10^{44}$ \\
        \hline
	\end{tabular}
\end{table*}


\section{Discussion}\label{discuss}

\subsection{Dense and Diffuse {Gas Masses and Kinematics}}
 
\begin{figure}
\centering
\includegraphics[width=\columnwidth]{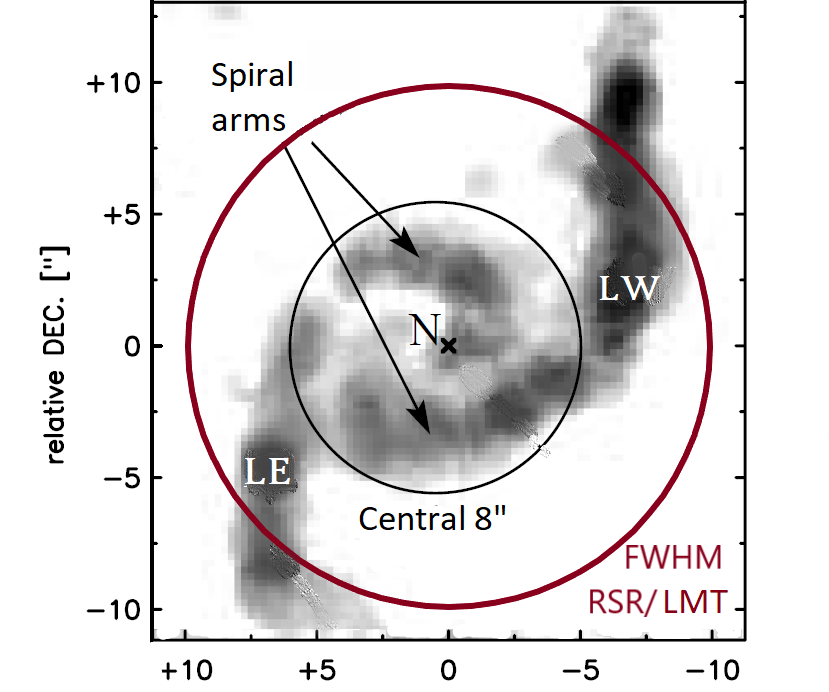}
\caption{A high spatial resolution ($\sim$150 pc) map of the center of NGC\,4303 observed with OVRO in the $^{12}$CO (1-0) line obtained by \citet{2002ApJ...575..826S} covering 22\arcsec$\times$22\arcsec. The center (black circle) has a diameter of 8\arcsec, and the red circle is RSR/LMT region of 20\arcsec which represents the LMT main beam for $^{13}$CO frequency. The bulk of molecular gas in this region obtained by \citet{2002ApJ...575..826S} was compared with that obtained in our study  {using} the $^{13}$CO line.} 
\label{MasaDisco}
\end{figure}

The molecular hydrogen mass obtained for the telescope's main beam center in line $^{13}$CO can be compared with that found by \citet{2002ApJ...575..826S} in a high spatial resolution study ($\sim 150$ pc) of gas traced by $^{12}$CO (J=1-0) line made with the \textit{Owens Valley Radio Observatory} (OVRO). As can  be seen in Fig.~\ref{MasaDisco},  {the molecular gas emission is found in two lanes: the western gas lane (LW) and the eastern gas lane (LE), both curving towards the nucleus (N)}. It was found that the central 8\arcsec \,  ($\sim 630$ pc) concentrates a mass $M(H_2)\,\sim\,$7$\times10^7$ M$_{\odot}$.  {Their estimation  used the conversion factor {$\alpha_{CO}$= $4.36$ M$_{\odot} $pc$^{-2}$(K km s$^{-1})^{-1}$} from \citet{1988A&A...207....1S}, with a $\pm$30$\%$ uncertainty}. In addition, {the southern component of LW and the north component of LE} that extend to an area of $\sim$22\arcsec$\times$22\arcsec, concentrate almost twice the value of the  nuclear disk or central 8$\arcsec$ mass,{ see Table 3 in \citet{2002ApJ...575..826S}.} {Consequently this yields} a total  {central} molecular mass $M(H_2)\sim$ 1.81$\times10^8$ M$_{\odot}$. 
If we use  {instead} the integrated intensity of the $^{12}$CO (J=1-0) line obtained from  literature  {\citep{2020A&A...635A.131I}}, the  molecular mass associated with this line derived from {$\Sigma_{mol}= 220\pm 79 $} M$_{\odot}$ pc$^{-2}$  {(see  \S 4.1)}, for an area with a diameter of 22${\arcsec}$ ($\sim$1.8~kpc) is $(5.7\pm \, 2.0)\times10^8$ M$_{\odot}$.  {Using the same conversion factor $\alpha_{CO}$ in both calculations, the factor of 3 difference in both masses could be due to the spatial resolution with which the observations were made, a beam of $\sim$150 pc vs. $\sim$1.8 kpc. In the second case, assuming a constant beam filling factor overestimates the amount of emission within the observed area.}

{In this study, the calculated molecular $H_2$ mass was  $M(H_2)$=(1.75$\pm$0.32)$\times 10^{8}$ M$_{\odot}$, which is in good agreement with the value reported by \citet{2002ApJ...575..826S}. We note that although the RSR/LMT observed region (the circle of 20\arcsec $\sim$1.6~kpc of diameter) does not completely cover the arms as shown in Fig.~\ref{MasaDisco}, most of the molecular gas is concentrated in the central region,  {which explains the similar values.}} Schinnerer’s work covers a slightly larger area than the one presented here,  {so to properly compare $M(H_2)$, we consider the value from  \citet{2002ApJ...575..826S} presented above as an upper limit.}


As we saw in the previous section, { the $M(H_2)$ value obtained with $^{12}$CO and Eq.~\eqref{DenSmolecular} is three times} the value obtained with $^{13}$CO. This difference may be due to the conversion factor $\alpha_{CO}$, as although in many studies it is assumed to be a constant value across various scales and environments, recent studies have found variations of 1-2 orders of magnitude that depend on the physical properties of the medium. \citet{Teng_2023} observed the central $\sim2$ kpc region of two barred spiral galaxies  at $\sim$100 pc resolution with $^{12}$CO and found a correlation between $\alpha_{CO}$ and the optical depth, an anticorrelation with kinetic temperature of the gas and low values of $\alpha_{CO}$ by factors of 4-15 in the central regions of these galaxies. As we can see, this variation has a significant impact on the calculation of molecular masses and the quantities derived from it, so care must be taken in choosing the value of $\alpha_{CO}$ according to the scale and characteristics of the ISM.  {In this paper, we consider the value of $M(H_2)$ obtained with $^{13}$CO to be the most accurate.}

The surface density of dense molecular gas ($\Sigma_{dense}$) 
 obtained  {using} Eq.~\eqref{DenSdenso} is integrated in a region where the molecular gas is not homogeneously distributed and thus $\Sigma_{dense}$ may be underestimated 
 (see Fig.~\ref{MasaDisco}). An important result is  the calculation of the amount of dense gas traced by the HCN in the same region $M_{dense}=(4.7\pm0.3) \times 10^{7}$ M$_{\odot}$.  {This value resulted} $\sim$4 times smaller that the total mass of molecular hydrogen. {This dense} 
 star-forming gas provide the conditions for the 200--250\,pc diameter circumnuclear ring studied by \citet{2016MNRAS.461.4192R}, through observations in the UV-optical-IR, where they reveal intense bursts of star formation. {{The relationship between  dense and molecular} mass is $M_{dense}$=0.21 $M(H_2$) {which is comparable to} $f_{dense}$= 0.13$\pm$0.06,{ within the uncertainties.} 


{Regarding gas kinematics, we analyzed the velocity of the molecular emission lines and compared them with those in previous studies}. The rotational curve of the galaxy traced by molecular gas $^{12}$CO J=(2-1) with a resolution of $\sim$150 pc from ALMA observations \citep[{Figure D3},][]{2020ApJ...897..122L}  shows the fitting of the observational data, resulting in asymptotic rotation velocity   {$V_0$\,=\,178.2$^{+75.1}_{-43.0}$\,km\,s$^{-1}$}. Meanwhile, the curves obtained by simulations of central bar formation made by  \citet{2022MNRAS.510.3899I}, show that, for a  {$r\leq 2$} kpc radius, the rotational velocity V$_{rot}$ from  the simulations (IsoB IC, blue curves in Figure A1 in that paper) are in the range of 150 to 175\,km\,s$^{-1}$.
 
 {In this} study, the velocities for the diffuse gas, i.e., $^{13}$CO and C$^{18}$O, gave us a mean velocity of $V_{CO}=$ 151$\,\pm\,$29\,km\,s$^{-1}$ that matches with the rotational velocity in  {the  {$r\leq 2$} kpc radii} obtained by \citet{2022MNRAS.510.3899I}, i.e., which covers  {partially the bar radius} $r_{bar}$\,=\,2.50\,$\pm$\,0.63 kpc \citep[e.g.,][]{2021MNRAS.500.2380Z},  {which is important because in this region the bar dominates the gas kinematics.}  {Meanwhile,} the mean value  of the velocity for dense gas tracers HCN, HNC, and HCO$^{+}$ is $V_{dense}$\,=\,178\,$\pm$\,60 
km~s$^{-1}$, which is consistent with the results obtained by \citet{2020ApJ...897..122L}. The dense gas velocity is  {comparable with the diffuse gas even having different FOVs, this is maybe due to the dominance of the structural component of the bar.}

\subsection{Emission line ratios}

The molecular emission line  ratio method helps quantify the molecular gas in specific regions of the galaxy, as in the case of AGN, and can provide information to characterize  the star formation or SMBH accretion processes.  {In this work, we are trying to determine if the presence of the AGN is hinted by the line ratios and to characterize the ISM from this central 1.6  {kpc} region.}

Previous studies have attempted to classify the activity of galaxies using molecular emission line diagrams  such as \,HNC/HCN  or highest values of \, HCN/HCO$^{+}$ \citep{2001ASPC..249..672K,2011A&A...528A..30C,2011MNRAS.418.1753J}.

{Another example is the use of the HNC/HCN ratio as an indicator of shock and non-standard heating, or the CS/HCN ratio, which should be greater than 1 at high column densities for both XDRs and PDRs, 
 suggesting a correlation with column density \citep{2008A&A...477..747B}}.

\begin{figure*}
\centering
\includegraphics[width=\textwidth]{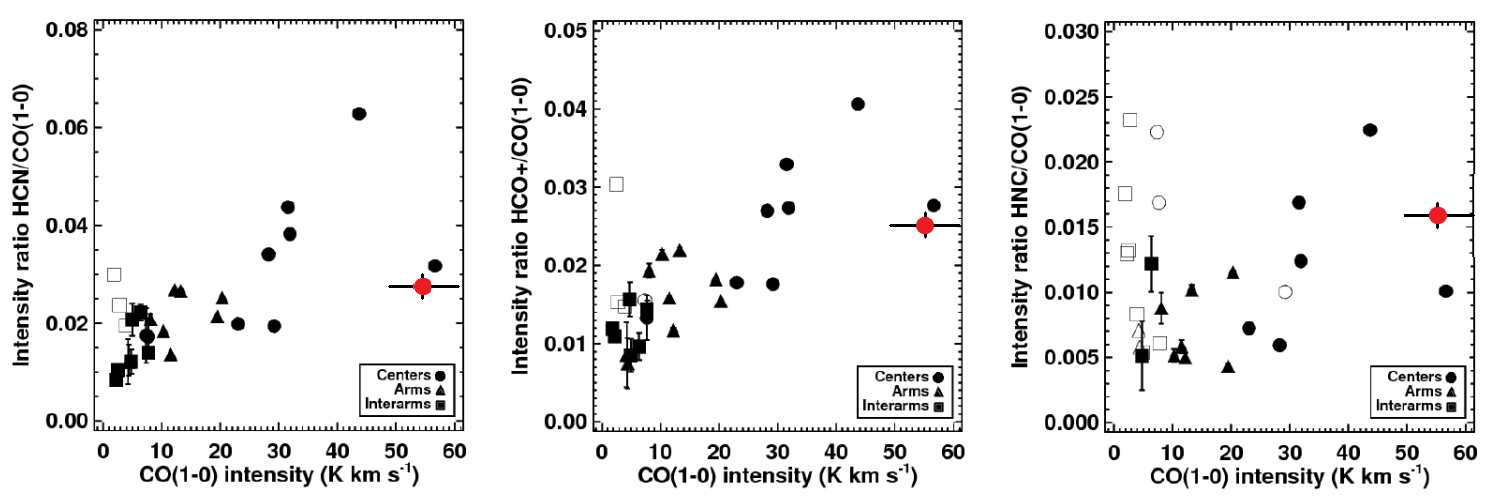}
\caption{Diagnostic diagrams left-to-right HCN/$^{12}$CO, HCO$^+$/$^{12}$CO and HNC/$^{12}$CO for a sample of nine nearby galaxies  {from}  {\citet{2019ApJ...880..127J}}. Squares are the values in the interarms regions, 
triangles are the values in the arms and circles are the values of the centers of these galaxies. Open symbols represent upper limits. The red circles represent the values for the central region of NGC\,4303 obtained in this study, which  is consistent with the location of the centers of other  spiral galaxies on the diagram.}
\label{Donaire}
\end{figure*}

\begin{figure*}
\centering
\includegraphics[width=\textwidth]{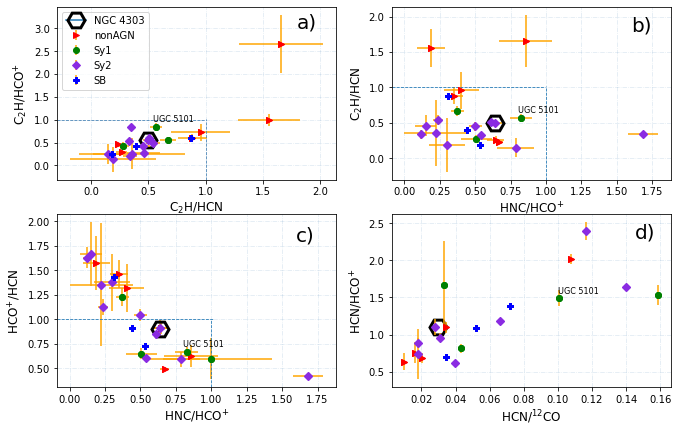}
\caption{Line ratio diagrams for a set of  {22  galaxies in the literature plus central 1.6 kpc of NGC\,4303}. Left to right: a) C$_2$H/HCN vs. C$_2$H/HCO$^+$, b) HNC/HCO$^+$ vs. C$_2$H/HCN, c) HNC/HCO$^+$ vs. HCO$^+$/HCN,  d) HCN/$^{12}$CO vs. HCN/HCO$^+$.  {The line ratio values from the literature are integrated values, the open black hexagons} are the values for the central 1.6 kpc of NGC\,4303. In addition, the values for UGC\,5101 obtained by \citet{2020MNRAS.499.2042C} in a similar study are labeled in each diagram.  {The dashed blue line enclosed is the region where the ratio is less than one.}}
\label{fig:coc}
\end{figure*}

{\citet{2019ApJ...880..127J} studied dense gas in nine nearby galaxies {at $\sim$1-2 kpc resolution} and showed that the main emission line ratios had a maximum value at the center {(with radius $\approx0.8$ kpc)} and decreased radially towards the outermost parts of the disk or spiral arms}. Using values from  {\citet{2019ApJ...880..127J}} and our study, we present in
Fig.~\ref{Donaire}    the diagnostic diagram of HNC, HCN, HCO$^{+}$ relative to $^{12}$CO. We observed an increasing relationship that clearly separated the galaxy regions, with the galactic centers being the most intense. The ratios HCN/$^{12}$CO, HCO$^+$/$^{12}$CO and HNC/$^{12}$CO presented in Table~\ref{tab:data2} for the center of NGC\,4303 are located in the diagnostic diagram  {as mentioned previously. As shown, our values are located in the region where the galactic central regions are also located}. It should be noted that the center of the galaxy closest to our point is the barred spiral galaxy NGC\,6946, which is also classified as a Sy 2.

\begin{figure*}
\centering
\includegraphics[width=11.2cm]{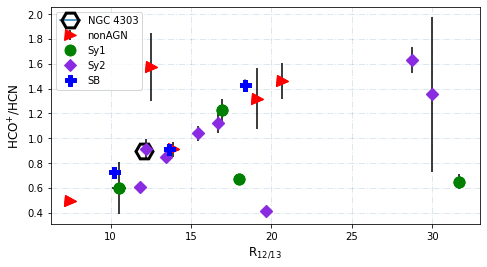}
\caption{R$_{12/13}$ vs. HCO$^{+}$/HCN for a sample of 12 nearby galaxies.  {The open black hexagon} is the value for the center of NGC\,4303. Although a  clear linear dependence can be seen between the lines, there is no dependence on galactic activity.}
\label{fig1213}
\end{figure*}

The ratio between $^{12}$CO and $^{13}$CO is denoted by $R_{12/13}$ and  has been studied for a large sample of galaxies because these lines are generally more intense and easier to detect \citep[e.g.,][]{2015yCat..35790101A, 2020A&A...635A.131I}. The value $R_{12/13}$ for the nucleus of the galaxy NGC\,4303 is 12.1\,$\pm$\,1.4,  {consistent with} the mean value $R_{12/13}$\,=\,13\,$\pm$\,6 of nearby starburst galaxies found by \citet{1995A&A...300..369A}.  This ratio provides information on the variation in optical depth in each galaxy, 
as well as the relative abundances of $^{12}$C and $^{13}$C. In the case of $^{12}$C, this can be produced by supernovas of massive stars, whereas $^{13}$C can be produced in low-mass stars or ion exchange $^{13}$C$^{+}$ in regions with temperatures $\sim10$~K   \citep[see, for example,][]{1998ApJ...494L.107K, 2014MNRAS.445.2378D}.

The HNC/HCN\,=\,0.58\,$\pm\,$0.04 ratio shows that HNC abundance is lower than that of HCN. {\citet{1998ApJ...503..717H} observed 19 nearby dark cloud cores with Nobeyama 45 m}. According to them, the HNC/HCN  value is a consequence of the temperature of the central molecular region $T_{ex}\,>\,$24~K, where  neutral-neutral reactions occur. In this manner, the HNC molecule is destroyed to form its isomer, HCN. This supports the  {use of $T_{ex}$\,=\,30 K to derive  surface densities with the corresponding value of $M(H_2$).}

{HCN/HCO$^{+}$} is the ratio most sensitive to density changes,  because the critical density of HCO$^{+}$ is one order  {of magnitude} smaller than that of HCN. Therefore, HCO$^{+}$ tends to recombine faster than HCN, when the medium is dense owing to free electrons. However,  {both molecules have similar abundances}, indicating that at high densities there is possibly  a mechanism that enhances the abundance of HCO$^{+}$. Studies of molecular abundance  by \citet{2005A&A...436..397M}  suggests that, in PDRs, this molecule abundance enhancement occurs at densities $n>10^{5}$ cm$^{-3}$. {As mentioned previously,  HCN/HCO$^{+}$ has also been used to determine  the nature of the nuclear-emitting source. In particular, values of HCN/HCO$^{+}$} greater than one, were found in XDRs, caused by  AGN with low density $n>10^{4}$ cm$^{-3}$ and $N(H_2)\,<\,$10$^{22}$ cm$^{-2}$ \citep[for example, ][]{2008A&A...488L...5L},  {which} agrees with the values of HCN/HCO$^{+}=1.10$ and $N(H_2$) found in our study  {(see Table~\ref{NM})}.    {Recently, \citet{2024arXiv240215436N} found in a 23 local ULIRGs sample than this ratio (J=3-2 lines) have no trend with AGN luminosity or implication in SFR, but the presence of outflows or inflows have the most crucial influence.} The HCN/HCO$^{+}$ ratio indicates a very similar abundance between the two molecules. Thus, HCO$^{+}$ can form in the dense region, that is, when $n_{critic}>10^{5}$ cm$^{-3}$, ionization must occur in the medium through shock waves or PDRs produced by young stars. The stellar populations of the circumnuclear region described by \citet{2019MNRAS.482.4437D} are young ($t<2$ Gyr) and add a high density to the medium, reinforcing the idea that the gas in the region has high ionization probabilities for radical formation.

Recently, \citet{2023MNRAS.521.3348N} presented the results of  {the} {ALMOND  (ACA Large-sample Mapping Of Nearby galaxies in Dense gas) survey,} where they traced the molecular gas density { using  resolved measurements of HCN(1-0)} across 25 nearby spiral galaxies, one of which was NGC\,4303. The HCN beam in that work (20$\arcsec$ $\sim$1.6 kpc) is similar to ours, and therefore, the results can be compared. The HCN integrated intensity in the 2 kpc galactocentric radius, had a peak temperature of 9.4 mK,   {similar to our value (see Table \ref{table:molecules})}.
 {In the upper left panel of Figure~6 of \citet{2023MNRAS.521.3348N}, the relationship between  HCN(1-0)/$^{12}$CO(2-1) and $(\Sigma_{mol})$ is presented, our logarithmic values are -1.55 and 2.05, respectively, which lie in their linear relation.}  {This data point is within the range of  values for $\log$ HCN/$^{12}$CO reported for NGC\,4303 (see their figure I4). Our $\Sigma_{mol}$ is on the lower end of the range reported in \citet{2023MNRAS.521.3348N}, but we note their value was calculated as a cloud-scale property at 150 pc resolution  using $\alpha_{CO}$, whose uncertainty can be attributed to variations in the galactocentric radius \citep[e.g.,][]{2013ApJ...777....5S}.}
 
 
{Diagnostic diagrams were constructed to determine possible relationships between type of galaxy activity and  emission lines. In Table~\ref{tab:muestra} we present a sample of 23 normal main sequence, starburst (SB), AGN {Sy 1} and AGN {Sy 2} galaxies with integrated intensities of HCN, HNC, C$_2$H, HCO$^{+}$, $^{12}$CO and $^{13}$CO reported in the literature. The sample is presented in Figure~\ref{fig:coc}, and} includes the values for the central region of NGC\,4303 as a black circle. In these diagrams no relationship between the  {line intensity ratios} and  galaxy activity is observed, the SB types and AGNs are mixed, and NGC\,4303 lies between the groups. Linearity trends can be observed between the ratios: C$_2$H/HCN vs. C$_2$H/HCO$^+$,  HNC/HCO$^+$ vs. HCO$^+$/HCN  {and} HCN/$^{12}$CO vs. HCN/HCO$^+$.

In the first relation, \textcolor{blue}{(a)} in Fig.~\ref{fig:coc}, we can see  a linear relation between these ratios, also the majority of the galaxies lie in the region where the ratios are less than one, with the especial cases of NGC\,3556 and IC~180. The diagram \textcolor{blue}{(d)} HCN/$^{12}$CO vs. HCN/HCO$^+$ shows an increasing relation  {too}, where, for values of HCN/HCO$^+$ less than one, most of the galaxies have a low fraction of dense gas  {($\lesssim 0.05$)} and for values greater than one the amount of dense gas, with respect to diffuse gas grows by a factor of 5-8. This last point may indicate a strong relationship between HCO$^+$ and CO. Additionally, having a dense gas fraction greater than 5$\%$ may be linked to the recombination of HCO$^+$ and increase the abundance of  HCN molecules.

Regarding the presence of C$_2$H in the central region may be associated with reactions produced by PDRs in massive, hot star formation regions  \citep[for example,][]{2015A&A...573A.116M}.{ This molecule, like other simple hydrocarbons ($^{13}$CN, CH, c-C$_3$-H$_2$) are the product of keeping a large amount of carbon ionized with warm gas associated with ultraviolet radiation, therefore, the intensity of the field is proportional to its abundance \citep[][]{2005ApJ...618..259M, Meier_2015}. It is worth noting that we have possible detections of some hydrocarbons that reinforce the idea of the presence of these PDRs.}

In Figure~\ref{fig1213}  we show R$_{12/13}$ vs. HCO$^{+}$/HCN and can observe another increasing relation, where  {the galaxies with values of HCO$^{+}$/HCN less than 1 present values of $R_{12/13}< 15$, that is, close to the mean value for nearby starburst  galaxies found by \citet{1995A&A...300..369A}. Note that this ratio is an indicator of the optical depth of the gas, that is, when $R_{12/13}$ increases, the medium becomes optically thicker. Therefore, HCO$^{+}$/HCN could be a possible indicator of highly obscured galaxies.}

As described above, the  {integrated} molecular gas diagrams cannot offer  a diagnosis of the presence or absence of obscured AGNs, as the {Sy 1} and {Sy 2} galaxies show no differences. However, they can tell us more about the abundance of the molecular gas in terms of the change in density and temperature conditions in the regions near the galactic centers. This can lead to characterization of the conditions of active star formation or its suppression in the regions surrounding an SMBH.

 {Separating the contribution of the central AGN from the rest of the galaxy is a problem. Observationally,} the telescope's main beam is usually larger than the galaxy diameter; hence, the intensity of the lines usually includes  contribution of arms (disk)  {and} the bar (if  {any}). In addition to this spatial resolution problem, it is important to detect high transitions of dense gas tracers with the disadvantage that they are even weaker in intensity than those with rotational transition J=(1-0), {when they are optically thin. An example of such a study is reported in \citet{2023ApJ...954..148I}}, a study with ALMA $\lesssim$0.5 kpc resolution,  where they detected the J=2-1, J=3-2 and J=4-3 high transition of several emission lines in 11 ULIRGs and showed the spatial variations of physical and chemical properties of molecular gas within  $\lesssim$2 kpc nuclear regions.}

\subsection{$L_{HCN}$ vs. $L_{IR}$ Relation}

{One of the outcomes of the SED fitting analysis is the total infrared luminosity $L_{IR}$, which is estimated as the sum of torus luminosity $L_{TORUS}$, and the dust luminosity, $L_{DUST}$. From the Bayesian analysis values we obtained $\log(L_{IR})$ = 10.96$\pm$0.04 $L_{\odot}$ (see Table 7).  {\citet{2003AJ....126.1607S} reported a similar value   $\log(L_{IR})$= 10.51 $L_{\odot}$  for NGC\,4303, determined using the fluxes in all four IRAS bands.} Both $L_{IR}$ values were larger than those of \citet{2020PASJ...72...41L}, { who reported} $\log(L_{IR})$= 9.6$\pm$0.03 $L_{\odot}$. However, \citet{2020PASJ...72...41L} processedcalibrated IR image data from Spitzer MIPS and Herschel PACS instruments and corrected to IRAM telescope's main beam { (which is close in size to the main beam of the LMT)}, to estimate $L_{IR}$.
For that reason, we use the value of $\log(L_{IR})$ from \citet{2020PASJ...72...41L}, together with our derived value for $\log(L_{HCN})$\,=\,6.87$\pm$0.08\, K km s$^{-1}$ pc$^{2}$ (Table~\ref{tab:lum}) to place NGC\,4303 in  figure 2 of \citet{2016ApJ...822L..26B} and figure 5 of \citet{2023MNRAS.521.3348N}. In both cases, NGC\,4303 lies close to the star formation sequence of a sample of nearby galaxies and individual molecular cloud cores; in the group of active star-forming disk galaxies. As we can see, the $L_{HCN}$ vs. $L_{IR}$ relation presented by \citet{2023MNRAS.521.3348N} is  independent of disc, arm or interarms regions in spatially resolved galaxies.

From a similar analysis for the galaxy UGC~5101 presented by \citet{2020MNRAS.499.2042C}, which was observed in the same season with the RSR/LTM, we obtain
$\log(L_{HCN})$\,=\,8.87$\,\pm\,$0.02\, K\,km \,s$^{-1}$\,pc$^{2}$ and taking $\log(L_{IR}$)\,=\,11.81\,$\pm$\,0.08\, $L_{\odot}$ from \citet{1989AJ.....98..766S}, this object falls among the Luminous Infrared Galaxies (LIRGs) in the correlation from \citet{2016ApJ...822L..26B}.  }

\subsection{AGN component and $SFR$ in NGC\,4303 from SED fitting}

 {According to the results of  SED fitting (see Table~\ref{tab:agn_results}), the AGN component 
in NGC\,4303 is rather low, with an AGN fraction value (\texttt{fracAGN}) of 0.2. 
Indeed, other studies that use CIGALE to fit AGN components to SEDs of galaxies have analyzed its 
reliability in determining the AGN contribution compared with optical classification and color-color
 diagrams \citep[for example,][]{Wang2020} and found a minimum value of \texttt{fracAGN} = 0.2 as a reliable 
lower limit for classifying AGNs using this method.
Thus, the value of \texttt{fracAGN} that we find, allows us to classify 
NGC\,4303 at least as marginally AGN, and could be in agreement with the difficulties 
to classify this galaxy as centrally dominated by star formation or a SMBH, and perhaps is speaking in 
favor of the fading phase AGN scenario \citep{2020ApJ...905...29E}.}

{To asses the goodness of the fit and validate the properties derived from it, 
we look at the value of the reduced $\chi^2$, which in our case is 3.1. 
We consider that this value corresponds to a fairly good fit and confirm this by 
visual inspection (see Fig.~\ref{fig:sed_fit}). We notice however, that the residuals at 
far-infrared wavelengths are higher than those at both, shorter and longer (submm) wavelengths,
but the model fluxes (red dots) still describe the shape of the observations. 
This could be caused by the fact that the observation at 100$\mu$m, just before the region of 
the higher residuals, is very well described by the model (residual $\sim$\,0), hence pinning 
the model at that point and preventing it from describing the far-infrared observations in a 
better manner.}

{Regarding the SFR in NGC\,4303, previous works have reported its value estimated using
UV and infrared maps.
\citet{2018ApJ...861L..18U} estimated the SFR from the combined GALEX FUV map, corrected for 
galactic extinction, and the WISE W4 infrared map. The maps were convolved to have an angular 
resolution of 15\arcsec\ and background subtracted. After this, they applied the prescriptions 
described in \citet{2012ARA&A..50..531K} and \citet{2013AJ....145....6J}, obtaining a 
$SFR=$5.24 M$_{\odot}$ yr$^{-1}$. A similar analysis was done by \citet{2020MNRAS.493.2872C} 
in a FOV of roughly the size of the galaxy (see their Fig.~1), obtaining a $SFR=4.37\pm 0.87$
M$_{\odot}$ yr$^{-1}$. \citet{2019ApJS..244...24L} calibrated the above mentioned prescriptions 
to match the results from \citet{2016ApJS..227....2S,2018ApJ...859...11S} who combined GALEX,
WISE and SDSS data to estimate the integrated SFR for low-redshift galaxies using CIGALE. 
The value they found is $SFR$=5.49$\pm$1.58  M$_{\odot}$ yr$^{-1}$. More recently, 
\citet{2021ApJS..257...43L} adopted the prescription of \citet{2019ApJS..244...24L} using a 
linear combination of FUV and W4 light, converting the luminosity of each band to SFR using 
the respective conversion factors and adding the UV and infrared terms to obtain the value 
$SFR$=5.37 M$_{\odot}$ yr$^{-1}$.}

{In summary, previous values of $SFR$ are similar to our CIGALE value, $SFR$=6.0$\pm$0.3, but most authors quote somewhat smaller values. We note that our analysis has the advantage that it incorporates UV to mm wavelength photometry, and the clumpy torus component. 
Assuming that the values of SFR from the literature are the true values, our similar result
validates the good quality of our CIGALE fit and the derived properties. Furthermore, using $SFR$ and $M_\star$ from \citep{2021ApJS..257...43L} we note that NGC\,4303 lies along the star-formation main sequence \citep[see for example,][for MaNGA SDSS-IV]{2019MNRAS.488.3929C}, for nearby galaxies.}





\section{Conclusions}

With the aim of studying the molecular gas in the obscured core of the galaxy NGC\,4303, in this work we analyze the emission line spectrum obtained with the \textit{Redshift Search Receiver} at the Large Millimeter Telescope (in its initial phase of 32 m) in the band of 3 mm (73 - 110 GHz), which correspond to 1.6 kpc centered in the $^{13}$CO (110.20 GHz) line.

We detected six molecular lines with S/N$>$3 in the J=(1-0) rotational transition: C$_2$H (87.31 GHz), HCN (88.63 GHz), HCO$^{+}$ (89.18 GHz), HNC (90.66 GHz), C$^{18}$O (109.78 GHz) and $^{13}$CO (110.20 GHz). The first four lines are gas with critical densities $n_{critic} > 10^{4}$ cm$^{-3}$ which makes them ideal for the study of  dense gas that is directly responsible for star formation in the galaxy. The average velocities obtained from the Gaussian fit were $151\pm29$ km s$^{-1}$ for the diffuse gas, and $178\pm60$ km s$^{-1}$ for the dense gas. The result for the diffuse gas velocity is in agreement with the result of \citet{2022MNRAS.510.3899I}, as well as in the observed $^{12}$CO J=(2-1) rotation curve of \citet{2020ApJ...897..122L}.

The isotopic varieties of carbon monoxide C$^{18}$O and $^{13}$CO trace the diffuse molecular gas with $n_{critic}\sim$ 10$^{3}$\,cm$^{-3}$ and taking advantage of the fact that the $^{13}$CO line is optically thin, that is, $\tau_{13} = 0.09\,\pm\,0.01$, the column densities relative to $^{12}$CO and H$_2$ for three excitation temperatures (10, 20 and 30 K) 
 were calculated. The results obtained with $T_{ex}$\,=\,30 K are $N(H_2)$\,=\,(5.14$\,\pm\,$0.94)$\times 10^{21}$\,cm$^{-2}$ and $M(H_2)$\,=\,(1.75$\,\pm\,$0.32)$\times10^{8}$\,M$_\odot$  {are in good agreement with the expected theory and  results obtained by other methods}. For example, the last result for the mass of molecular hydrogen is of the same order of magnitude as that found for a central region of 22\arcsec$\times$22\arcsec using a $^{12}$CO(1-0) high resolution map obtained by \citet{2002ApJ...575..826S}. Using the HCN  {integrated} intensity it was possible to calculate the amount of dense gas in the observed area. The value $M_{dense}=(4.7\pm0.3) \times 10^{7}$ M$_{\odot}$ is only 4 times smaller than the amount of gas in H$_2$. 



The intensities of the emission lines of the dense and diffuse molecular gas tracers clearly show that the abundance with which these species enrich the chemistry of the medium is due to certain star formation activity in the circumnuclear region. PDRs are a consequence of the interaction of young stars or regions of star formation and are responsible for generating this abundance of radicals in a very dense region that would cause them to recombine quickly. The accretion mechanisms of this gas towards the center responsible for such density must be linked to both the spiral arms and the bar that reach the {circumnuclear region}, which may be influenced by the gravitational potential of the SMBH. {Meanwhile, when comparing the  line ratios obtained here} with those of
a sample of galaxies in the literature, no clear trend is observed to distinguish nuclear activity {(SF or AGN)} from normal galaxies.

Our spectral energy distribution  analysis using CIGALE code shows that NGC\,4303 has a large clumpy dusty torus, with the following parameters obtained based on 
the Bayesian analysis: aperture above the equatorial plane 
\texttt{sigma} = 56$^\circ \pm$9$^\circ$, dust optical depth 
\texttt{tau\_V}=184$\pm$38, number of dust clouds along the equatorial plane \texttt{N\_0}=14$\pm$2, an inner \, to outer radius \, ratio 
of \texttt{Y\_ratio}=97$\pm$12, and a viewing angle of \texttt{incl}$\sim$70$^{\circ}$. By considering the classical unified AGN model 
\citep{1993ARA&A..31..473A,1995PASP..107..803U}; this viewing angle is in agreement with the Sy 2 
classification previously given by \citet{2006A&A...455..773V}.

Furthermore, for NGC\,4303 CIGALE yields the total AGN and torus luminosities of $L_{\rm AGN}$ = (1.5$\,\pm\,$0.6) $\times 10^{44}$ erg s$^{-1}$, $L_{\rm TORUS}$ = (7.1\,$\pm$\,2.8)$\times 10^{43}$ erg s$^{-1}$; while the galaxy dust 
luminosity is $L_{\rm DUST}$ =  (2.8$\,\pm$\,0.1) $\times 10^{44}$ erg s$^{-1}${, which yield a total infrared luminosity  $L_{\rm IR}$ =  (3.51$\,\pm$\,0.30) $\times 10^{44}$ erg s$^{-1}$.
Finally, the star formation rate obtained  is  $SFR=$ 6.0\,$\pm$\,0.3 M$_\sun$/yr, which together with the total stellar mass positions this galaxy along the star-formation main-sequence for normal nearby galaxies.}

{From the molecular gas and dusty torus analysis, we conclude that the central 1.6 kpc 
emission from NGC\,4303 is a mixture of an AGN with a marginal contribution of $\leq$20\%, most probably a Type 2, with a large clumpy dusty torus and a starburst host galaxy, 
as evidenced by intense dense molecular gas lines (C$_2$H, HCN, HCO$^{+}$, and HNC). We found that dense
gas contributed significantly to the total molecular gas mass.}

\bigskip\bigskip




{\bf Acknowledgements.} The authors thank the referee for a detailed and thorough review of the manuscript and for her/his valuable comments, which have enriched our work and are greatly appreciated. AS acknowledges support from graduate studies scholarship from CONAHCYT and CONAHCYT/SNI research assistant fellowship. ICG, EB, MHE and AS acknowledge financial support from DGAPA-UNAM grant IN-119123 and CONAHCYT grant CF-2023-G-100. 
MHE acknowledges financial support from CONAHCYT program {\it Estancias Posdoctorales por M\'exico}. We acknowledge support from CONAHCYT-Mexico, during the construction and Early ~Science Phase of the ~Large ~Millimeter ~Telescope~~ Alfonso Serrano (LMT/GTM), as well as support from the US National Science Foundation via the University Radio Observatory program, the Instituto Nacional de Astrof\'isica, \'Optica y Electr\'onica (INAOE), and the University of Massachusetts (UMASS). We also thank  LMT/GTM Observatory technical staff during the observations and data processing periods. This research made use of the SIMBAD database, operated at CDS, Strasbourg, France, and the NASA/IPAC Extragalactic Database (NED), which is operated by the Jet Propulsion Laboratory, California Institute of Technology,
under contract with the National Aeronautics and Space Administration.
\nolinenumbers

\begin{table*}[!t]
 \caption{Molecular line transitions detected in NGC\,4303 and other Galaxies
}
\label{tab:muestra}
\begin{center}
  \begin{tabular}{cccccccc}
\hline\hline \\
Galaxy & $\mathrm{\dfrac{C_2H}{HCO^+}}$ & $\mathrm{\dfrac{C_2H}{HCN}}$ & $\mathrm{\dfrac{HCN}{HCO^+}}$ & $\mathrm{\dfrac{HNC}{HCO^{+}}}$ & $\mathrm{\dfrac{HCN}{^{12}CO}}$ & Activity$^*$& Reference$^{**}$\\
\\\hline \hline
{  NGC\,4303}  &  0.55 & 0.50 & 1.10  &  0.64   & 0.03  & Sy2 & a\\
{  UGC\,5101}  & 0.84 & 0.56 &  1.50 & 0.83  & 0.10 & Sy1.5& b\\
IC 180  & 2.65  & 1.66  & 1.60   & 0.86  & - & - & c\\
NGC\,1614  & 0.26 & 0.54 & 0.74   &   0.22   & 0.02 & Sy2 & c\\
NGC\,3079   & 0.48 & 0.54& 0.89   &  0.24    & 0.02  & Sy2 & c\\
NGC\,4194  & 0.73 & 0.96 &  0.76 &  0.40    & 0.02  & - & c\\
NGC\,4388   & 0.14  & 0.19 & 0.73  &   0.30  & - & Sy2 & c\\
NGC\,4418    & 0.25  & 0.15  & 1.69   &  0.79   & - & Sy2 & c\\
NGC\,6090    & 0.27  & 0.46  & 0.60   &   0.15  & - & Sy2 & c\\
NGC\,6240    & 0.21  & 0.34  & 0.61   & 0.12   & 0.04  & Sy2 & c\\
NGC\,7771    & 0.29  & 0.26  & 1.10   & 0.65  & 0.03  & -  & c\\
NGC\,660    & 0.43  & 0.45  & 0.96   & 0.50 & 0.03 & Sy2 & c\\
NGC\,3556    & 0.99  & 1.56  & 0.64   &  0.19   & 0.01 & - & d \\
NGC\,2273   & -  & -  & 1.67   &  1.00  & 0.03  & Sy1-2 & d\\
NGC\,5236   & 0.43  & 0.39  & 1.09   & 0.44    & 0.05 & SB & e\\
NGC\,253   & 0.60  & 0.51  & 1.18  &  0.61   & 0.07 & Sy2 & e\\
M82   & 0.61  & 0.87  & 0.70   &0.31    & 0.03 & SB & e \\
M51   & 0.47  & 0.23  & 2.02  &  0.67   & 0.11 & - & e\\
NGC\,1068 &  0.54  & 0.33  & 1.64   &  0.54  & 0.14  & Sy2 & e\\
NGC\,7469   & 0.55  & 0.67  & 0.82 &  0.37   & 0.04 & Sy1.2 & e\\
ARP 220   & 0.83  & 0.35  & 2.39  &  1.68 & 0.12 & Sy2 & e\\
MRK 231 & 0.43 & 0.28 & 1.54   &  0.51  & 0.16 & Sy1 & e\\
IC 342    & 0.25  & 0.18  & 1.38  & 0.54   & 0.07 & SB & f \\ 
\hline\hline 
\end{tabular}
\begin{minipage} {0.9\textwidth}%
\begin{center}
{Notes:}   $^{*}$Activity type from NASA/IPAC Extragalactic Database (NED).\\
~~~~$^{**}$Dense gas emission line ratios from: $^a$ This work, $^b$ \citet{2020MNRAS.499.2042C}, $^c$ \\\citet{2011MNRAS.418.1753J}, $^d$ \citet{2011A&A...528A..30C},  $^e$ 
\citet{2015A&A...579A.101A}, $^f$ \\
\citet{2018PASJ...70....7N}. \\ 
\end{center}
  \end{minipage}
\end{center}
\label{table8}
\end{table*}

\bibliography{bib,CTAGN}
\end{document}